\documentclass[preprint,aps,showpacs,nofootinbib]{revtex4}

\usepackage{graphicx}
\usepackage{dcolumn}
\usepackage{bm}
\usepackage{axodraw}

\newcommand{\be}{\begin{equation}}      
\newcommand{\ee}{\end{equation}}        
\newcommand{\bear}{\begin{eqnarray}}    
\newcommand{\eear}{\end{eqnarray}}      
\newcommand{\beqstar}{\begin{eqnarray*}}        
\newcommand{\eeqstar}{\end{eqnarray*}}

\newcommand{\tr}{{\rm tr}}

\newcommand{\ov}[1]{\overline{#1}}

\begin{document}

\preprint{EFI-02-73, UFIFT-HEP-02-6, BUHEP-02-20}

\title{Radiative Corrections to Kaluza-Klein Masses \vskip.4in }

\author{Hsin-Chia Cheng}
 \email{hcheng@theory.uchicago.edu}
 \affiliation{Enrico Fermi Institute, The University of Chicago, Chicago,
   IL 60637, USA}
\author{Konstantin T. Matchev}%
 \email{matchev@phys.ufl.edu}
\affiliation{Department of Physics, University of Florida, Gainesville, FL 32611, USA
}%
\author{Martin Schmaltz}
 \email{schmaltz@bu.edu}
\affiliation{Physics Department, Boston University, 
Boston, MA 02215, USA
\vskip 1cm
}%


\begin{abstract}
Extra-dimensional theories contain a number of
almost degenerate states at each Kaluza-Klein level. If extra
dimensional momentum is at least approximately conserved then
the phenomenology of such nearly degenerate states depends
crucially on the mass splittings between KK modes.
We calculate the complete one-loop radiative
corrections to KK masses in general 5 and 6 dimensional
theories. We apply our formulae to the example of universal
extra dimensions and show that the radiative corrections are essential
to any meaningful study of the phenomenology.
Our calculations demonstrate that Feynman diagrams with
loops wrapping the extra dimensions are well-defined
and cut-off independent even though higher dimensional
theories are not renormalizable.
\end{abstract}

\maketitle

\section{\label{sec:level1}Introduction}

Radiative corrections are known to play an important role for precision
measurements, but are generally not expected to radically change
the nature of high energy ``discovery'' processes like the
production and decay of new particles in collider experiments.

In this paper we point out that this expectation can be completely
wrong with respect to the collider physics of some extra-dimensional
models.
Radiative corrections are crucial for determining the decays
of Kaluza-Klein (KK) excitations. This is because at tree
level KK masses are quantized, and all momentum
preserving decays are exactly at threshold.
Radiative corrections then become the dominant effect
in determining which decay channels are open.

Consider for example the simplest case of a massless field
propagating in a single circular extra
dimension with radius $R$. This theory is equivalently described
by a four dimensional theory with a tower of states with
tree level masses $m_n=n/R$.
The integer $n$ corresponds to the quantized momentum $p_5$ in the
compact dimension and becomes a quantum number (KK number) under a $U(1)$
symmetry in the 4d description. The tree level
dispersion relation of a 5d-massless particle is fixed by
Lorentz invariance of the tree level Lagrangian
\be
E^2=\vec p^2 + p_5^2 = \vec p^2 + m_n^2\ ,
\label{eq:dispersion}
\ee
where $\vec p$ is the momentum in the usual three spatial directions.
Ignoring branes and orbifold fixed points, KK number
is a good quantum number and is preserved in all interactions and
decays. We see from eq.~(\ref{eq:dispersion}) that at tree
level the KK modes of level $n>1$ are exactly at threshold for
decaying to lower level KK modes.
For example, in 5d QED with massless photons and electrons the reaction
\be
e^{(2)} \rightarrow e^{(1)} + \gamma^{(1)}
\label{eq:e2decay}
\ee
is exactly marginal at tree level.
It is straightforward to include electroweak symmetry breaking masses.
This gives no mass shift to the photon and its KK modes and
generates masses $\sqrt{m_n^2+m_e^2}$ for the electrons at KK level $n$.
Including these shifts one finds that the reaction
(\ref{eq:e2decay}) is just barely forbidden by phase space,
and one concludes that all electron KK modes are stable. However,
using realistic values $m_e\sim$ MeV and $R^{-1}\sim$ TeV, 
the difference between the total masses on both sides of equation
(\ref{eq:e2decay}) normalized to the KK mode masses is only of
order $m_e^2/m_n^2\sim 10^{-12}$.
Clearly, this minuscule mass splitting is completely irrelevant if
there are radiative corrections to eq.~(\ref{eq:dispersion}) which
would start at order $\alpha \sim 10^{-2}$. This is reminiscent of
the case of wino-LSP in supersymmetric models where the tiny tree
level wino mass splitting is overwhelmed by the radiative corrections
\cite{Cheng:1998hc}.

We now show that there are indeed radiative corrections to the
KK-masses.
The dispersion relation (\ref{eq:dispersion}) follows from
local 5d Lorentz invariance of the tree level Lagrangian. However,
5d-Lorentz invariance is broken by the compactification. This
breaking is non-local and cannot be seen
in the renormalized couplings of the local 5d Lagrangian,
but it contributes to the 4d masses of KK modes because of their
delocalized wave functions in the fifth dimension.
More explicitly, the leading mass corrections
$\delta m_n^2$ to eq.~(\ref{eq:dispersion}) come from loop
diagrams with internal
propagators which wrap around the compactified dimension.
The sign and $n$-dependence of these corrections
determines which decay channels are open and which KK modes
are stable. For the example of 5d-QED, we find radiative
corrections at order $\alpha$ as anticipated; they render the
reaction (\ref{eq:e2decay}) allowed with phase space of order
$\alpha R^{-1} \sim 10$ GeV. 

In this paper we compute mass corrections at one loop for a general theory
with fields of spin $0,\frac12$ and $1$. Our results are
finite and well defined.  At first sight, this might seem surprising
since the 5d theory is not renormalizable. However, the 5d Lorentz
violating corrections to KK mode masses involve propagation over
finite distances (around the extra dimension) and are exponentially
suppressed for momenta which are large compared to the
compactification scale. Thus our results are UV-finite and
do not depend on the
choice of regulator as long as it is 5d Lorentz invariant
and sufficiently local.

Applying these results to the Standard Model requires introducing
an additional complication. Obtaining chiral fermions
in 4d from a 5d theory is only possible with additional breaking
of 5d Lorentz-invariance. Two frequently discussed choices are
introducing chiral fermions on branes or imposing
orbifold boundary conditions on fermions in the bulk.
We focus on the latter because we wish to minimize the breaking of
5d Lorentz invariance. The resulting model in which all the Standard Model
fields live in the bulk of an orbifold
is known as ``Universal Extra Dimensions'' \cite{Appelquist:2001nn}.
We consider the orbifolds $S^1/Z_2$ and $T^2/Z_2$.

Both orbifolds have fixed points which break extra-dimensional
translation invariance, and we expect
new interactions localized on the fixed points.
Clearly, the presence of such localized interactions
violates 5d momentum conservation, and KK number
is no longer preserved. However, a discrete subgroup remains
unbroken. In the $S^1/Z_2$ case, this is ``KK-parity'',
a parity flip of the extra dimension. In the 4d description
KK-parity is a $Z_2$ symmetry
under which only KK-modes with odd KK-number are charged. The symmetry
implies that the lightest KK particle at level 1
(the LKP) is stable. Note that KK-parity and the LKP play
an analogous role to R-parity and the LSP in supersymmetry.

In the presence of orbifold boundaries higher level
KK-modes can decay to lower level KK-modes via
KK number violating interactions. These decays compete with
KK number preserving decays, and it becomes a phenomenologically
important question which channels dominate. The answer can be
understood very simply. Since the KK number violating interactions
exist only on the boundaries they turn into volume suppressed
couplings between KK modes. This implies that even though KK number
violating decays have larger phase space they are more strongly
suppressed because they are proportional to the square of smaller
coupling constants.
Therefore, the question of which momentum preserving decays 
are allowed by phase space remains phenomenologically important
also in theories on orbifolds.

In addition to giving rise to new interactions,
the boundary terms also include 5d Lorentz violating
kinetic terms which contribute to the masses of
KK modes and are important in determining
decay patterns. 
In reference \cite{Georgi:2001ks} it was shown that the coefficients of
boundary terms receive logarithmically divergent contributions
at one loop. Thus it is not only possible to include boundary terms
in orbifold theories, it is inconsistent not to include them.
The coefficients of these terms correspond to new
parameters of the theory. They contain incalculable contributions
from unknown physics at the cutoff as well as contributions
from loops in the low-energy theory which we compute in this
paper.

This paper is structured as follows. In the next section
we compute radiative corrections to masses of KK modes for
scalars, fermions, and gauge fields in
a 5d theory on a circle. In Section 3 we discuss the additional
complications which arise for orbifolds and compute
the renormalization of boundary couplings. In Section 4
we apply the results of the previous sections to the Standard
Model in ``Universal Extra Dimensions'' and determine the
complete one-loop corrected spectrum. Section 5 contains our
conclusions. Details of our calculations can be found in 
Appendices.

\section{Bulk corrections from compactification}

To begin, we discuss the simplest higher dimensional
theory: an extra dimension compactified on
a circle $S^1$ with radius $R$ ($x_5+2\pi R \sim x_5$). We assume that
5d Lorentz invariance is respected by the short-distance
physics, and is only broken by the compactification. The momentum in the 5th 
dimension, which is quantized in units of $1/R$, becomes a mass for
the KK modes after compactification. If 5-dimensional Lorentz 
invariance were exact, the KK 
mode masses coming from the 5th dimensional momentum would not receive
corrections. For example,
the kinetic term of a scalar field living in 5 dimensions is 
\be
{\cal L} \supset Z \partial_\mu \phi\, \partial^\mu \phi 
- Z_5 \partial_5 \phi\, \partial_5 \phi, \quad
\mu=0,1,2,3\ .
\ee
Both $Z$ and $Z_5$ receive divergent quantum corrections. However, if 
5-dimensional Lorentz invariance were exact, these contributions
would be equal, so that the masses of the KK modes coming from the 
$(\partial_5 \phi)^2$ term would stay uncorrected.
More generally, exact Lorentz invariance would imply that the energy is only
a function of $|\vec{p}|^2+ p_5^2$, and hence $E^2=|\vec{p}|^2+ p_5^2 +m^2$
does not receive $p_5$-dependent corrections. All KK mode masses
would be given by $p_5^2+m^2$ with the same $p_5$-dependence, and
the only correction would be due to renormalization of the
zero mode mass $m$.

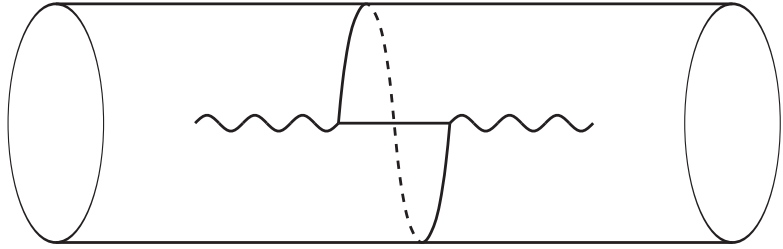
\begin{figure}[tb]
\begin{center}
{
\unitlength=1.5 pt
\SetScale{1.5}
\SetWidth{0.7}      
\normalsize    
{} \qquad\allowbreak
\begin{picture}(200,100)(0,0)
\Oval(15,50)(30,12)(0)
\Oval(185,50)(30,12)(0)
\Line(86,50)(114,50)
\Photon(114,50)(150,50){2}{3}
\Photon(86,50)(50,50){2}{3}
\Line(15,80)(185,80)
\Line(15,20)(185,20)
\DashCurve{(93,80)(95,78)(97.5,70)(99,60)(100,50)(101,40)(102.5,30)(105,22)(107,20)}{2}
\Curve{(93,80)(91,78)(88.5,70)(87,60)(86,50)}
\Curve{(107,20)(109,22)(111.5,30)(113,40)(114,50)}
\end{picture}
}
\caption{\label{fig:winding} Lorentz violating loop.}
\end{center}
\end{figure}

However, 5-dimensional Lorentz invariance is broken at long distances
by the compactification, so in general the masses of the KK modes do receive
radiative corrections. Feynman diagrams are sensitive to the Lorentz
symmetry breaking if they have an internal loop which winds around
the circle of the compactified dimension,
as shown in Fig.~\ref{fig:winding}, so that it can tell that
this direction is different from the others. 
%
%
This is a non-local effect
as the size of the loop can not be shrunk to zero. Such
non-local loop diagrams are well-defined and finite, even though the
higher-dimensional theory is non-renormalizable.

We can isolate the finite 5d Lorentz violating corrections from
the divergent 5d Lorentz invariant corrections by employing a very
simple subtraction prescription: from every loop diagram in the
compactified theory we subtract the corresponding diagram of
the uncompactified theory. The UV divergences are canceled
because the two theories are identical at short distances, but the
KK mass corrections are unaltered because the subtraction is
5d Lorentz invariant.

To make this more explicit, first note that momenta in the compact
dimension are discrete. Therefore the five-dimensional phase space
integral
\be
\int \frac{d^5 k}{(2\pi)^5} \cdots 
\label{5dphasespace}
\ee
becomes
\be
\frac1{2\pi R} \sum_{k_5} \int \frac{d^4 k}{(2\pi)^4} \cdots 
\label{kksum}
\ee
for compact dimensions.

Our subtraction prescription is to simply subtract
eq.~(\ref{5dphasespace}) from eq.~(\ref{kksum}) for each diagram.
To better understand the physical meaning of this prescription we rewrite
the KK-sum using the Poisson resummation identity
\be
\frac1{2\pi R} \sum_{m=-\infty}^{\infty} F(m/R) =
               \sum_{n=-\infty}^{\infty} f(2\pi R\, n),
\label{Poisson}
\ee
where $f(x)$ and $F(k)$ are related by Fourier transformation
\be
f(x) = {\cal F}^{-1}\left\{ F(k)\right\}=\int_{-\infty}^\infty
\frac{dk}{2\pi}\, e^{-ikx} F(k)\ .
\ee
The resummation formula turns a sum over KK numbers $m$
(or KK momenta $k_5=m/R$) into a sum over winding numbers $n$
(or position space windings $n 2\pi R$ around the fifth dimension).
Note that the
$n=0$ term in the sum is identical to the phase space integral
of an uncompactified extra dimension
\be
f(0) = \int_{-\infty}^\infty \frac{dk_5}{2\pi}\, F(k_5)\ 
= \int \frac{d^5 k}{(2\pi)^5}\ \cdots\  .
\ee
Thus our subtraction prescription simply amounts to leaving
out the divergent $n=0$ term in the re-summed expression for
each Feynman diagram. The remaining terms in the sum (with $n \ne 0$)
correspond to particle loops with net winding $n$ around
the compactified dimension.%
\footnote{More precisely, they correspond to diagrams in which
the internal propagators form a non-contractible loop around
the extra dimension. The parameter $n$ is the winding number
of the internal loop. The diagrams with a contractible loop are
5d Lorentz invariant and get subtracted.}
They are all finite and so is their sum.

\begin{figure}[tb]
\begin{center}
{
\unitlength=1.5 pt
\SetScale{1.5}
\SetWidth{0.7}      
\normalsize    
{} \qquad\allowbreak
\begin{picture}(140,100)(0,0)
\Photon(5.0,50.0)(40.0,50.0){2}{3}
\Photon(100.0,50.0)(135.0,50.0){2}{3}
\ArrowArc(70,50)(30,0,180)
\ArrowArc(70,50)(30,180,360)
\Text(70,10)[]{$k,k_5$}
\Text(70,90)[]{$k-p,k_5-p_5$}
\Text(20,60)[]{$p,p_5$}
\end{picture}
}
\caption{\label{fig:vac-pol} Vacuum polarization diagram.}
\end{center}
\end{figure}
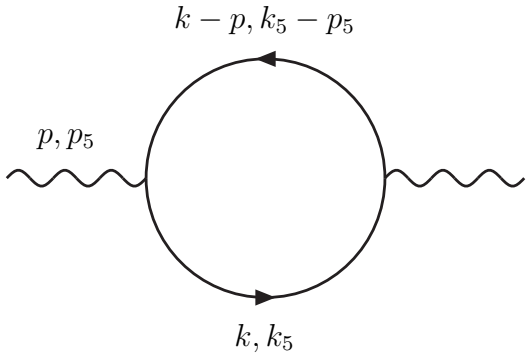

To illustrate the calculation, we consider the relatively
simple example of QED in 4+1 dimensions with one spatial dimension 
compactified on a circle. We will calculate
the correction to the masses of KK photons due to the electron loop.
The one loop vacuum polarization (Fig.~\ref{fig:vac-pol})
is given by
\bear 
i\Pi_{\mu\nu}&=& -e^2 \sum_{k_5} \int \frac{d^4 k}{(2\pi)^4}
\tr \Bigg[\gamma_\mu \frac{1}{\not{k}+i\gamma_5 k_5} 
\gamma_\nu \frac{1}{(\not{k}-\not{p})+i\gamma_5(k_5-p_5)}
\Bigg] \\
&=&
-4\,e^2 \sum_{k_5} \int \frac{d^4 k}{(2\pi)^4}
\frac{k_\mu(k_\nu-p_\nu)+ k_\nu(k_\mu-p_\mu) - g_{\mu\nu} k(k-p)
+g_{\mu\nu} k_5(k_5-p_5)
}{(k^2-k_5^2)[(k-p)^2-(k_5-p_5)^2]} \nonumber
\eear
where $p, \, k$ are 4-momenta, $k_5=m/R$ with $m=$integers,
and the volume factor $1/(2\pi R)$ has been absorbed into the
gauge coupling $e^2=e_5^2/(2\pi R)$.
%
%

As usual we use Feynman parameterization to combine the denominators,
\be
i\Pi_{\mu\nu}= -4\,e^2 \int_0^1 d\alpha \sum_{k_5} 
\int \frac{d^4 k}{(2\pi)^4}  \frac{
  N_{\mu\nu}}{
[k^2-k_5^{\prime 2}+\alpha(1-\alpha)(p^2-p_5^2)]^2} 
\ee
where
\be
 N_{\mu\nu} = 2 k_\mu k_\nu +g_{\mu\nu}(-k^2 +
\alpha(1-\alpha)(p^2-p_5^2)
+(2\alpha-1) p_5 k'_5 + k^{\prime 2}_5)
-2\alpha(1-\alpha) p_\mu p_\nu ,
\ee
and $k'_5=k_5-\alpha p_5$.
To calculate the correction to the masses of the KK modes, we concentrate
on the terms proportional to $g_{\mu\nu}$,
\be
\Pi_{\mu\nu}=g_{\mu\nu}\, \Pi_1- p_\mu p_\nu \Pi_2.
\ee
We can set $p^2=p_5^2$ in the
leading order approximation.
Replacing $k_\mu k_\nu$ by $g_{\mu\nu}k^2/4$, 
and performing the Wick rotation, we have
\be
\Pi_1 = -4\, e^2 \int_0^1 d\alpha \sum_{k_5}
\int \frac{d^4 k_E}{(2\pi)^4} \frac{\frac{1}{2} k_E^2 + (2\alpha-1)p_5 k'_5
+k^{\prime 2}_5}{(k_E^2+ k^{\prime 2}_5)^2}.
\ee
It is convenient to rescale $k_E,\, k_5,\, k'_5,\, p_5$ 
by $1/R$ so that they become dimensionless numbers and 
$k_5,\,p_5$ run over integers. Using the formula
\be
\frac{1}{A^r}=\frac{1}{(r-1)!}\int_{0}^{\infty} d\ell \ell^{r-1}
e^{-A\ell},
\ee
we obtain
\be
\Pi_1 = -\frac{4\, e^2}{R^2}
\int_0^1 d\alpha \sum_{k_5} \int \frac{d^4 k_E}{(2\pi)^4}
\int_0^\infty d\ell \,\ell \left[\frac{1}{2} k_E^2 + (2\alpha-1)p_5 k'_5
+k^{\prime 2}_5\right] e^{-(k_E^2+ k^{\prime 2}_5)\ell}.
\ee
Next, we perform the $d^4 k_E$ integral
\bear
\Pi_1 &=& -\frac{4\, e^2}{16\pi^2 R^2} \int_0^1 d\alpha
\int_0^\infty d\ell\, \ell \sum_{k_5}
\left[\frac{1}{\ell^3} + \frac{(2\alpha-1)p_5 k'_5}{\ell^2}
+ \frac{k^{\prime 2}_5}{\ell^2} \right] e^{-k^{\prime 2}_5\ell}
\nonumber \\
&=& 
-\frac{ e^2}{4\pi^2 R^2} \int_0^1 d\alpha
\int_0^\infty dt \sum_{k_5}
\left[1+ \frac{(2\alpha-1)p_5 k'_5}{t}
+ \frac{k^{\prime 2}_5}{t} \right] e^{-k^{\prime 2}_5/t},
\eear
where $t=1/\ell$.
Now we use the Poisson resummation formula, eq.(\ref{Poisson}),
to turn the sum over $k_5$ into a sum over winding numbers.
The inverse Fourier transformations needed are
\bear
{\cal F}^{-1}\left\{ e^{-k_5^2/t} \right\}
&=& \sqrt{\frac{t}{4\pi}} e^{-x^2 t/4} \nonumber \\
{\cal F}^{-1}\left\{ k_5 e^{-k_5^2/t} \right\}
&=& -i\, \frac{xt}{2}\, \sqrt{\frac{t}{4\pi}} e^{-x^2 t/4} \nonumber \\
{\cal F}^{-1}\left\{ k_5^2 e^{-k_5^2/t} \right\}
&=& \left( -\frac{x^2 t^2}{4} +\frac{t}{2} \right) 
\sqrt{\frac{t}{4\pi}} e^{-x^2 t/4} \nonumber \\
{\cal F}^{-1} \left\{ F(k'_5=k_5-\alpha p_5)\right\}
&=& f(x) e^{-i\alpha x p_5} 
\eear
The result is
\bear
\Pi_1
&=& -\frac{e^2}{2\pi R^2} \sum_{x=2\pi n} \int_0^1 d\alpha \,
e^{-i\alpha x p_5} \int_0^\infty dt \sqrt{\frac{t}{4\pi}} e^{-x^2 t/4}
\left[ \frac{3}{2} - i\left( \alpha -\frac{1}{2}\right) x p_5
-\frac{x^2 t}{4} \right]
\nonumber \\
&=& -\frac{e^2}{2\pi R^2} \sum_{x=2\pi n} \int_0^1 d\alpha \,
e^{-i\alpha x p_5}
\left[ \frac{3}{2}\, \frac{2}{|x|^3} - 
i\left( \alpha -\frac{1}{2}\right) x p_5\, \frac{2}{|x|^3}
-\frac{x^2}{4}\, \frac{12}{|x|^5} \right]
\nonumber \\
&=& -\frac{e^2}{2\pi R^2} \sum_{n=-\infty}^{\infty}
\int_0^1 d\alpha \, e^{-i\alpha 2\pi n p_5} 
\left(-i(2\alpha-1) 2\pi n p_5\right) \frac{1}{|2\pi n|^3}.
\eear
For the zero mode ($p_5=0$), we have $\Pi_1=0$, i.e., there is
no correction to the mass as expected by gauge invariance.
For nonzero KK modes, the correction to their masses is obtained by
dropping the divergent $n=0$ term as discussed above
\be
\delta m_{KK}^2= -\frac{e^2}{2\pi R^2} \sum_{n \neq 0} \frac{2}{|2\pi n|^3}
= - \frac{e^2}{4\pi^4 R^2} \sum_{n=1}^{\infty} \frac{1}{n^3}
= - \frac{e^2 \zeta(3)}{4\pi^4 R^2},
\ee
which is finite and independent of the KK level.

It is straightforward to follow the same procedure to calculate the corrections
in a more general theory which contains non-Abelian gauge fields, fermions, and
scalars. In our calculation, we assumed that the zero mode masses
are much smaller than the compactification scale
so that we can ignore them in the calculations. With the possible exception
of the Higgs boson and the top quark, this is also the case of interest
for applications to the Standard Model.
(For non-vanishing zero mode mass $m_0 \ll 1/R$, there will be KK level
dependent corrections suppressed by $m_0^2/p_5^2$.)
The one-loop contributions from various diagrams are listed in 
Appendix~\ref{app:bulk} and we summarize the results here.

The correction to the KK mode masses for the gauge field is given by
\be
\delta m^2_{V_{KK}}=\frac{g^2 \zeta(3)}{16\pi^4 R^2} \left(3 C(G) +
\sum_{\rm real\; scalars} T(r_s) -4\sum_{\rm fermions} T(r_f)\right),
\label{gaugebulk}
\ee
where $C(G)\delta_{ab}=f_{acd}f_{bcd}\, (=N$ for 
$SU(N))$, and $T(r)\delta_{AB}=\tr(T^A T^B)$ 
is the Dynkin index of the representation $r$,
normalized to be 1/2 for the fundamental representation of $SU(N)$.
The sum over scalars is over the real components and needs to be multiplied
by 2 for a complex scalar. Note that for a supersymmetric
theory the correction vanishes as it has to because the KK
gauge bosons are BPS states.
As in the case of QED5, the zero mode mass is not corrected as
dictated by gauge invariance.

A similar calculation yields the correction to the mass of the
zero mode of $A_5$. We find 
\be
\delta m^2_{A_5^0} = 3\; \delta m^2_{V_{KK}}\ ,
\ee
which is in agreement with earlier calculations \cite{Hosotani}.
Note that the KK modes of $A_5$ are ``eaten'' and become longitudinal
components of the KK gauge fields.

For fermions, we find
\be
\delta m_{f_{KK}}=0\ .
\ee

Fine tuning is required for a scalar to be light, as its 
(Lorentz invariant) mass receives power divergent corrections no matter
whether the extra dimension is compact or not. We are interested in
the difference between the corrections to the KK modes and the zero mode,
assuming that the zero mode mass has been fine tuned to be smaller
than the compactification scale. In calculating
the potentially 5d Lorentz violating contributions from loops with nonzero 
winding number, we find that the lowest order
corrections to the squared masses of the zero mode and KK modes are
the same,
\be
\delta m^2_{S_{KK}} = \delta m^2_{S_0} \ .
\ee
Therefore, they can be absorbed into the (infinitely renormalized)
zero mode mass,
and the $n$-th KK mode mass is simply given by
\be
m^2_{S_n}= \frac{n^2}{R^2} + m_0^2
\ee
with no corrections at the lowest order.

In the above calculations we have ignored graviton loops
\cite{Contino:2001nj}. Their effects on KK mass splittings
are negligible as they are suppressed by powers of $M_{Pl} R$.

\section{Orbifold compactifications}

In the previous section we considered the simplest compactification on
a circle (the generalization to a torus in more extra dimensions is
discussed in Appendix~\ref{app:bulk}). However, a higher
dimensional fermion has 4 or more components. Its four dimensional 
zero mode consists of both left-handed and right-handed fermions
when compactified on a torus, and the resulting four dimensional
theory is vector-like. To obtain chiral fermions in four dimensions,
we need more complicated compactifications. One possibility
is to compactify the extra dimensions on an orbifold.
In this section, we consider
the simplest example, an $S^1/Z_2$ orbifold, where $Z_2$ is the 
reflection symmetry $x_5 \to -x_5$. In addition to their indirect
transformation via their $x_5$-dependence, fields can be even or odd
under this $Z_2$ symmetry. A consistent assignment is to
have $A_\mu,\, \mu=0,1,2,3$ even, and $A_5$ odd for the gauge field,
and $\psi_L$ even (odd), $\psi_R$ odd (even) for the fermions.
The scalars can be either even or odd. From a field theory point
of view, the orbifold is simply a line segment of length $L=\pi R$
with boundary points (orbifold fixed points) at $x_5=0,\, \pi R$.
Even (odd) fields have Neumann (Dirichlet) boundary conditions,
$\partial_5 \phi =0\, (\phi =0)$ at $x_5=0,\, \pi R$.

The KK decomposition for even and odd fields is given by
\bear
\Phi_+ (x, x_5) &=& \frac{1}{\sqrt{\pi R}}\, \phi^{(0)}_+ (x) +
\sqrt{\frac{2}{\pi R}} \,\sum_{n=1}^{\infty} \cos \frac{nx_5}{R}\:
\phi^{(n)}_+ (x) , \nonumber \\
\Phi_- (x, x_5) &=& 
\sqrt{\frac{2}{\pi R}} \,\sum_{n=1}^{\infty} \sin \frac{nx_5}{R}\:
\phi^{(n)}_- (x) .
\label{kkdecomp}
\eear
The zero mode of the odd field is projected out by the orbifold
$Z_2$ symmetry (or Dirichlet boundary conditions). For a fermion $\psi$,
only $\psi_L$ (or $\psi_R$) has a zero mode, hence we obtain
a chiral fermion in the four dimensional theory. Similarly, the
$A_5$ zero mode is projected out and there is no massless
adjoint scalar from the extra component of the gauge field.

The orbifold introduces additional breaking of higher dimensional
Lorentz invariance which
leads to further corrections to KK mode masses.
The orbifold fixed points break translational symmetry
in the $x_5$ direction, therefore
momentum in the $x_5$ direction (KK number) is no longer conserved,
and we expect mixing among KK modes. However, a translation by $\pi R$
in the $x_5$ direction remains a symmetry of the orbifold.
We can see from eq.(\ref{kkdecomp}) that under this transformation the 
even number ($n=$even) KK modes are invariant while the odd number
($n=$odd) KK modes change sign. Therefore, KK parity $(-1)^{KK}$ 
(not the $Z_2$ in $S^1/Z_2$) is still a good symmetry.
Note that KK-parity is a flip of the line segment about
it's center at $x_5=\pi R/2$ combined with the $Z_2$ transformation
which flips the sign of all odd fields.

Because 5d Lorentz and translation invariance are broken
at the orbifold boundaries, radiative corrections generate
additional Lagrangian terms which are localized at the boundaries
and don't respect 5d Lorentz symmetry.
The boundary terms contribute to masses and
mixing of KK modes. To calculate them, we follow 
the work by Georgi, Grant, and Hailu~\cite{Georgi:2001ks}.
Fields on the $S^1/Z_2$ orbifold can be
written as
\bear
\phi(x,\, x_5) &=& \frac{1}{2} \left( \Phi(x,\, x_5) \pm 
\Phi(x,\, -x_5)\right),\nonumber \\
\psi(x,\, x_5) &=& \frac{1}{2} \left( \Psi(x,\, x_5) \pm \gamma_5
\Psi(x,\, -x_5)\right),
\label{orbifold}
\eear
where $\Phi, \Psi$ are unconstrained 5-dimensional boson and fermion
fields, and the upper (lower) sign, $+(-)$, corresponds to $\phi, \psi_R$ 
being even (odd) under $x_5\to -x_5$.
The propagators such as 
\be
S(x-x',\, x_5-x'_5)=\langle \psi(x,\, x_5) \ov{\psi}(x',\, x'_5) \rangle
\ee
can be expressed in terms of unconstrained fields~(\ref{orbifold}).
The results are
\be
S(p, p_5, p'_5)=
\frac{i}{2} \left\{ \frac{\delta_{p_5,p'_5}}{\not{p}+i\gamma_5 p_5}
\mp \frac{\delta_{-p_5,p'_5}}{\not{p}+i\gamma_5 p_5}\gamma_5 \right\}
\ee
for the fermion,
\bear
D_{\mu\nu}(p, p_5, p'_5) &=& 
\frac{-ig_{\mu\nu}}{2}\left\{\frac{\delta_{p_5,p'_5} + \delta_{-p_5,p'_5}}{p^2-p_5^2}
\right\},
\nonumber \\
D_{55}(p, p_5, p'_5) &=&
\frac{-ig_{55}}{2}\left\{\frac{\delta_{p_5,p'_5} - \delta_{-p_5,p'_5}}{p^2-p_5^2}
\right\},
\eear
for the gauge field (in the Feynman-'t~Hooft gauge), and
\be
D(p, p_5, p'_5) =
\frac{i}{2}\left\{\frac{\delta_{p_5,p'_5}\pm \delta_{-p_5,p'_5}}{p^2-p_5^2}
\right\}
\ee
for the scalar boson. $p_5$ and $p'_5$ are the outgoing and incoming 
fifth dimensional momenta (KK numbers). They can be different because
5d momentum is not conserved.

We calculate the one-loop diagrams with these modified propagators.
%
%
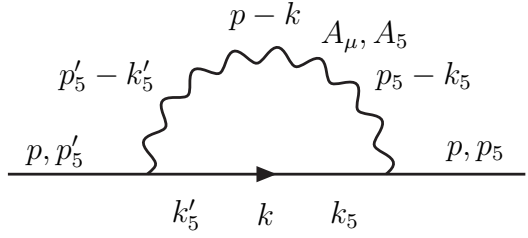
\begin{figure}[tb]
\begin{center}
{
\unitlength=1.5 pt
\SetScale{1.5}
\SetWidth{0.7}      
\normalsize    
\begin{picture}(140,100)(0,30)
\ArrowLine(5.0,50.0)(135.0,50.0)
\PhotonArc(70,50)(30,0,180){2}{9}
\Text(70,40)[]{$k$}
\Text(50,40)[]{$k'_5$}
\Text(90,40)[]{$k_5$}
\Text(70,90)[]{$p-k$}
\Text(30,75)[]{$p'_5-k'_5$}
\Text(110,75)[]{$p_5-k_5$}
\Text(17,56)[]{$p,p'_5$}
\Text(123,56)[]{$p,p_5$}
\Text(95,85)[]{$A_\mu,A_5$}
\end{picture}
}
\caption{\label{fig:electron} Electron self-energy diagram.}
\end{center}
\end{figure}
Consider, for example, the one-loop contribution to the
electron self energy in 5d QED (Fig.~\ref{fig:electron}). Let us first
focus on the summation over momenta in the fifth dimension.
The summations are of the form
\bear
&&\sum_{k_5,k'_5} \left(\delta_{k_5,k'_5}+\delta_{-k_5,k'_5}\gamma_5
\right) \left(\delta_{p_5-k_5,p'_5-k'_5}+\delta_{-(p_5-k_5),p'_5-k'_5}
\right) \nonumber \\
&=& 
\left(\delta_{p_5,p'_5}+\delta_{-p_5,p'_5}\gamma_5\right) \sum_{k_5} 
+ \sum_{k_5} \left(\delta_{2k_5,p_5+p'_5}+\delta_{2k_5,p_5-p'_5}\gamma_5\right) \ .
\eear
Up to a factor of $\frac12$, the term proportional to
$\delta_{p_5,p'_5}+\delta_{-p_5,p'_5}\gamma_5$ reproduces
the corresponding diagram in 5d QED on a circle, and
we can simply recycle the result of the previous section.
The relative factor of $\frac12$ arises because the $Z_2$ orbifolding
projects out half of the states of the theory on $S^1$.
The second term gives rise to new contributions to the
self energy which violate 5d momentum by integer multiples of
$2/R$. We will see shortly that these terms are log divergent.
The corresponding counter terms are localized on the fixed points
of the orbifold at $x_5=0$ and $x_5=\pi R$.

Denoting the ``boundary'' contribution to the self
energy by $\overline \Sigma(p;p_5,p'_5)$ we have
\bear
&-&\!\!\!i\overline\Sigma(p;p_5,p'_5) = \\
&-&\!\!\frac{g^2}{4} \sum_{k_5} \int \frac{d^4 k}{(2\pi)^4}
\Bigg[\frac{\gamma^\nu (\not{k} +i\gamma_5 k_5)\gamma^\mu g_{\mu\nu}
-\gamma_5 (\not{k}+i\gamma_5 k_5)\gamma_5}{
(k^2-k_5^2)[(k-p)^2-(k_5-p_5)^2]}\Bigg]
\bigg( \delta_{2k_5, p_5+p'_5} \pm\delta_{2k_5, p_5-p'_5}\gamma_5 \bigg)
\nonumber
\eear
where the first term in the numerator comes from the 4-dimensional
gauge field components and the second term comes from the 5th component
of the gauge field. After Feynman parametrization and Wick rotation,
this becomes
\bear
-i\overline{\Sigma}(p;p_5,p'_5) &=& \frac{ig^2}{4} \sum_{k_5} \int_0^1 d\alpha 
\int\frac{d^4 k_E}{(2\pi)^4}
\frac{(\alpha \not{p} +5i\gamma_5 k_5)
( \delta_{2k_5, p_5+p'_5} \pm\delta_{2k_5, p_5-p'_5}\gamma_5 )}
{[k_E^2 -\alpha (1-\alpha)p^2
+k_5^2-2\alpha k_5 p_5 -\alpha p_5^2 ]^2}
\nonumber \\
&\to& \frac{ig^2}{64\pi^2}\ln\frac{\Lambda^2}{\mu^2}
\sum_{k_5} \Bigg[\frac{1}{2} \not{p} + 5i\gamma_5 k_5 \Bigg] 
\bigg( \delta_{2k_5, p_5+p'_5} \pm\delta_{2k_5, p_5-p'_5}\gamma_5 \bigg)
\nonumber \\
&=& \frac{ig^2}{64\pi^2}\ln\frac{\Lambda^2}{\mu^2}
\Bigg[ \not{p} \frac{1\pm\gamma_5}{2} + 5i\gamma_5 p_5 \frac{1\pm\gamma_5}{2}
+5i \gamma_5 p'_5 \frac{1\mp\gamma_5}{2} \Bigg]
\left.^{\text{for even}}_{R(p_5-p'_5).}\right. 
\eear
The arrow in the second line indicates that we have only kept the
leading logarithmic divergence. In the log, $\Lambda$
represents the cutoff and $\mu$ is the renormalization scale.
The equality in the final line holds only for $R(p_5-p'_5)$ even,
for odd differences we have $\overline \Sigma (p,p_5,p'_5)=0$.

This result can be understood (following \cite{Georgi:2001ks}) as the
renormalization of terms in the 5d Lagrangian which are localized
at the boundaries of the orbifold. Fourier transforming to
position space, we obtain
\be
\delta \overline{{\cal L}} \supset \bigg(
\frac{\delta(x_5) +\delta(x_5-L)}{2}\bigg)
\frac{L g^2}{64\pi^2} \ln\frac{\Lambda^2}{\mu^2}
\Bigg[ \ov{\psi}_+ i \not{\!\partial} \psi_+ + 5 (\partial_5 \ov{\psi}_-)
\psi_+ + 5 \ov{\psi}_+ (\partial_5 \psi_-) \Bigg],
\label{boundary}
\ee
where $L$ appears because of a change in normalization of fields
in going from 4d to 5d; $L$ combines with the 4d gauge
coupling to give $L g^2 = g_5^2$. The delta functions are
normalized to $\int_0^\epsilon \delta(x)dx=1$.
We have been using Feynman gauge in the above calculation.
For general 't~Hooft $\xi$ gauges, one can show that the coefficients
in front of $i\!\!\!\!\not{\!\partial}$ and $\partial_5$ are given by
$1+2(\xi-1)$ and $5+(\xi-1)$, respectively.

The logarithmically divergent result means that we should include counter
terms localized at the boundaries to cancel the divergence. Our
calculation only determined the running contribution between
the cut-off $\Lambda$ and $\mu$, given initial values for the
boundary terms at $\Lambda$. We implicitly assumed in our calculations
that the boundary terms at the cut-off are small. If
large boundary terms were present, they would mix KK modes of different
levels and correspondingly shift their masses. Both effects would have to be taken
into account in calculating the radiative corrections.
The KK spectrum would then have a complicated dependence on the
unknown boundary terms at the high scale. We continue to
assume that there are no large boundary terms, and the logarithmic
divergences can be absorbed into the cutoff $\Lambda$
with $\Lambda$ not too large. Note that this assumption is self-consistent
because the boundary terms which are generated by radiative
corrections are loop-suppressed.

The leading order correction to the mass of the $n$-th KK mode
is obtained from Lagrangian terms which are quadratic in the $n$-th KK mode.
Mass corrections due to the mixing among different KK modes are
of higher order.

We expand the boundary terms (\ref{boundary})
in terms of the KK modes and consider the modification of
the kinetic terms for the $n$-th KK mode, ($n\neq 0$),
\be
Z_{n+} \ov{\psi}_{n+} i \not{\!\partial} \psi_{n+} + 
\ov{\psi}_{n-} i \not{\!\partial} \psi_{n-}
+ Z_{n5} \left( \ov{\psi}_{n+} \partial_5 \psi_{n-} 
- \ov{\psi}_{n-} \partial_5 \psi_{n+}\right),
\ee
where
\bear
Z_{n+}& =& 1 +  2(1+2(\xi-1)) \frac{g^2}{64\pi^2} \ln \frac{\Lambda^2}{\mu^2},
\nonumber \\
Z_{n5} &=& 1+ 2 (5+(\xi-1))  \frac{g^2}{64\pi^2} \ln \frac{\Lambda^2}{\mu^2}.
\eear
Note that $Z_{n-}=1$ because $\psi_{n-}$ vanishes on the boundary.
After rescaling $\psi_{n+}$ to canonical kinetic terms, the correction
to the KK mode mass is given by 
\be
\frac{\bar{\delta} m_n}{m_n}= \frac{Z_{n5}}{\sqrt{Z_{n+}}}-1 
= \frac{9}{4}\, \frac{g^2}{16\pi^2}
\ln\frac{\Lambda^2}{\mu^2},
\ee
which is independent of the gauge parameter $\xi$. The correction
is proportional to the $n$-th mode mass $n/R$, in contrast with
the bulk contribution discussed in the previous section.

For a more general theory which contains non-Abelian gauge fields,
fermions and scalars, the radiatively generated boundary terms from
various diagrams are listed in Appendix~\ref{app:boundary}. In the following, we
summarize the one-loop  corrections to the KK mode masses. 
We always assume that the boundary terms are
small, and can be treated as perturbations.

The corrections to the masses of KK modes for gauge bosons, fermions,
$Z_2$ even scalars, and $Z_2$ odd scalars are given by 
\bear
\bar{\delta} m_{V_n}^2 &=& m_n^2\, 
\frac{g^2}{32\pi^2}\, \ln \frac{\Lambda^2}{\mu^2}
\left[ \frac{23}{3}\, C(G) -\frac{1}{3} 
\sum_{\rm real\; scalars}\bigg(T(r)_{\rm even} -T(r)_{\rm odd}\bigg)\right],
\\
\bar{\delta} m_{f_n} &=& m_n\, \frac{1}{64\pi^2} \ln\frac{\Lambda^2}{\mu^2}
\left[ 9\,C(r)\,g^2- \sum_{\rm even\, scalars} 3\, h_+^2
+\sum_{\rm odd\, scalars} 3\, h_-^2\right],
\\
\bar{\delta} m^2_{S_{+n}} &=& \overline{m}^2 + m_n^2 \frac{1}{32\pi^2}
\ln \frac{\Lambda^2}{\mu^2} \left[ 6g^2 T(r) - \sum_{\rm even\, scalars}
\frac{\lambda_{++}}{2} +\sum_{\rm odd\, scalars} \frac{\lambda_{+-}}{2}
\right],
\\
\bar{\delta} m^2_{S_{-n}} &=&  m_n^2 \frac{1}{32\pi^2}
\ln \frac{\Lambda^2}{\mu^2} \left[ 9g^2 T(r) + \sum_{\rm even\, scalars}
\frac{\lambda_{+-}}{2} -\sum_{\rm odd\, scalars} \frac{\lambda_{--}}{2}
\right],
\eear
where $h$ and $\lambda$ are Yukawa and quartic scalar couplings
respectively. Their normalization is chosen to yield vertices with
no numerical factors in the Feynman rules. The $\overline{m}^2$ in the
expression for the even scalars contains a contribution
$+2\overline{m}^2$ to the KK mode mass from a
boundary mass term, minus a contribution $\overline{m}^2$ to the
zero mode mass from the same boundary term. The relative factor of two
between zero mode and KK modes comes from the normalization of the
wave functions in eq.~(\ref{kkdecomp}).

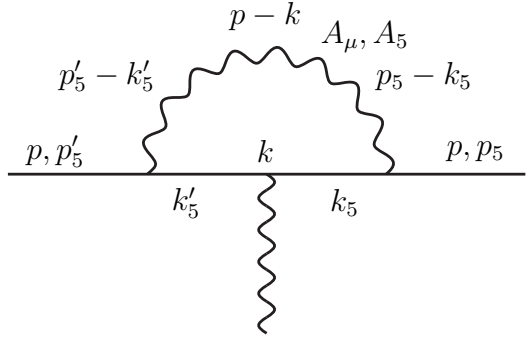
\begin{figure}[tb]
\begin{center}
{
\unitlength=1.5 pt
\SetScale{1.5}
\SetWidth{0.7}      
\normalsize    
\begin{picture}(140,100)(0,0)
\Line(5.0,50.0)(135.0,50.0)
\PhotonArc(70,50)(30,0,180){2}{9}
\Text(70,56)[]{$k$}
\Text(50,43)[]{$k'_5$}
\Text(90,43)[]{$k_5$}
\Text(70,90)[]{$p-k$}
\Text(30,75)[]{$p'_5-k'_5$}
\Text(110,75)[]{$p_5-k_5$}
\Text(17,56)[]{$p,p'_5$}
\Text(123,56)[]{$p,p_5$}
\Text(95,85)[]{$A_\mu,A_5$}
\Photon(70,50)(70,10){2}{5}
\end{picture}
}
\caption{\label{fig:vertex} One-loop diagram for the KK number violating
vertex in the 5 dimensional QED.}
\end{center}
\end{figure}

The boundary terms also induce KK number violating couplings. Because
KK parity is not broken, KK number can only be violated by even units
in these couplings. Using the QED on $S^1/Z_2$ example, we can calculate
the one-loop vertex diagram for the KK number violating coupling between
the photon and the electron, Fig.~\ref{fig:vertex}.
%
%
The result is simply to replace $\not{\!\partial}$ in eq.~(\ref{boundary})
by the covariant derivative $\not{\!\!D}$. To obtain the couplings among
the physical eigenstates, however, we have to take into account
the kinetic and mass mixing effects on the external legs.
A more detailed discussion is in Appendix~\ref{app:vertex}.
The result can be related to the mass corrections from the boundary terms
as both come from operators localized at the boundaries.
For example, we find that the 
$\overline{\psi}_{0} \gamma^\mu T^a P_+ \psi_0 A_{2\mu}$ coupling is given by
\be
 \frac{g}{\sqrt{2}} \left[
\frac{\bar{\delta}(m_{A_2}^2)}{m_{2}^2}- 2\,\frac{\bar{\delta} (m_{f_2})}{m_{2}}
\right].
\ee
On the other hand, couplings involving the zero mode gauge boson are
governed by gauge invariance which implies that KK number violating
interactions such as
$\overline{\psi}_{2} \gamma^\mu T^a P_+ \psi_0 A_{0\mu}$ vanish.

\section{The Standard Model in Universal Extra Dimensions}

We now apply the results obtained in the previous two sections to
the Standard Model in extra dimensions.
The KK modes of Standard Model fields receive additional tree level
mass contributions from electro-weak symmetry breaking which we have
not taken into account in the calculations of the previous sections.
Here, we include all these contributions but we ignore
effects which involve both electro-weak symmetry breaking and
radiative corrections. They are suppressed by both
$\frac{g^2}{16 \pi^2}$ and $\frac{v^2}{m_n^2}$ and are 
numerically negligible.

We consider the case in
which all the Standard Model fields propagate in the same extra dimensions,
(universal extra dimensions)~\cite{Arkani-Hamed:2000hv,Appelquist:2001nn}.
Theoretical motivations for considering such 
scenarios include electroweak symmetry breaking~\cite{Arkani-Hamed:2000hv},
the number of fermion generations~\cite{Dobrescu:2001ae}, 
proton stability~\cite{Appelquist:2001mj}. Here we take a
phenomenological approach and consider the simplest case
of one universal extra dimension compactified
on an $S^1/Z_2$ orbifold. The orbifold compactification is necessary
to produce chiral fermions in four dimensions. 
In ~\cite{Appelquist:2001nn,Agashe:2001xt,Appelquist:2001jz}
it was shown that the
current constraint on the compactification scale for one
universal extra dimension is only about 300 GeV.
Because of tree-level KK number conservation, KK states can only
contribute to precision observables in loops, and
direct searches for KK states require pair production.
If the compactification scale is really so low, then KK states
will be copiously produced at future colliders~\cite{Rizzo:2001sd,Macesanu:2002db}.
As we have argued in the introduction, the
radiative corrections have to be taken into account in any
meaningful study of the phenomenology of these KK modes.

We assume the minimal field content of the Standard Model in one
extra dimension. The fermions $Q_i,\, u_i,\, d_i,\, L_i,\, e_i,\, i=1,\,2,\,3$
are all 4-component fermions in 4+1 dimensions. (The upper case letters
represent $SU(2)$ doublets and the lower case letter represent $SU(2)$
singlets.)
Under the $Z_2$ orbifold symmetry,
$Q_L,\, u_R,\, d_R,\, L_L,\, e_R$ are even so that they have zero
modes, which are identified with the Standard Model fermions.
Fermions with opposite chirality ($Q_R,\, u_L,\, d_L,\, L_R,\, e_L$)
are odd and their zero modes
are projected out. In order to allow Yukawa couplings
the Higgs field must be even under the $Z_2$.

To obtain the corrections to the masses of the KK modes of the Standard
Model fields we simply substitute into the formulae from
the previous two chapters and include appropriate group theory and
multiplicity factors. 
The bulk corrections are given by (bulk contributions in the
$S^1/Z_2$ orbifold are half of those in the $S^1$ compactification) 
\bear
\delta\, (m_{B_n}^2) &=&  -\frac{39}{2}\, \frac{g'^2\,\zeta(3)}{16\pi^4}  
\left(\frac{1}{R}\right)^2,
\nonumber \\
\delta\, (m_{W_n}^2) &=&   -\frac{5}{2}\,\frac{g_2^2\,\zeta(3)}{16\pi^4}  
\left(\frac{1}{R}\right)^2,
\nonumber \\
\delta\, (m_{g_n}^2) &=& -\frac{3}{2} \, \frac{g_3^2\,\zeta(3)}{16\pi^4}  
\left(\frac{1}{R}\right)^2,
\nonumber \\
\delta( m_{f_n}) &=& 0,
\nonumber \\
\delta(m_{H_n}^2) &=& 0,
\label{delta0}
\eear
where $B_n$ are the KK modes of the $U(1)$ hypercharge gauge boson,
$W_n$ are the KK modes of the $SU(2)_W$ gauge bosons and
$g_n$ are the KK modes of the gluon.

The boundary terms receive divergent contributions which require
counter terms. The finite parts of these counter terms are
undetermined and remain as free parameters of the theory.\footnote{This 
is reminiscent of the case of low energy supersymmetry,
where in the absence of an explicit theory of supersymmetry breaking
we do not know the values of the soft masses at high scales.
Nevertheless, we can compute their renormalization within
a given visible sector model like the MSSM. Hence one can
predict the superpartner masses only under specific assumptions
about their values at the high scale.} Here we shall
make the simplifying assumption that the boundary kinetic terms vanish 
at the cutoff scale $\Lambda$ and compute their renormalization
to the lower energy scale $\mu$.
The corrections from the boundary terms are then given by
\bear
\bar{\delta}\, m_{Q_n} &=& m_n\,\left(3\, \frac{g_3^2}{16\pi^2}
+ \frac{27}{16}\,\frac{g_2^2}{16\pi^2} + \frac{1}{16}\,\frac{g'^2}{16\pi^2}
\right) \,\ln\frac{\Lambda^2}{\mu^2},
\nonumber \\
\bar{\delta}\, m_{u_n} &=& m_n\,\left(3\, \frac{g_3^2}{16\pi^2}
  + \frac{g'^2}{16\pi^2}
\right) \, \ln\frac{\Lambda^2}{\mu^2},
\nonumber \\
\bar{\delta}\, m_{d_n} &=& m_n\,\left(3\, \frac{g_3^2}{16\pi^2}
  + \frac{1}{4}\,\frac{g'^2}{16\pi^2}
\right) \, \ln\frac{\Lambda^2}{\mu^2},  \nonumber \\
\bar{\delta}\, m_{L_n} &=& m_n\,\left(
 \frac{27}{16}\,\frac{g_2^2}{16\pi^2} + \frac{9}{16}\,\frac{g'^2}{16\pi^2}
\right) \, \ln\frac{\Lambda^2}{\mu^2},  \nonumber \\
\bar{\delta}\, m_{e_n} &=& m_n\,
 \frac{9}{4}\,\frac{g'^2}{16\pi^2}
\, \ln\frac{\Lambda^2}{\mu^2}, \nonumber \\
\bar{\delta}\, (m_{B_n}^2) &=& m_n^2\, \left(-\frac{1}{6}\right)\, 
\frac{g'^2}{16\pi^2}
\ln\frac{\Lambda^2}{\mu^2}, \nonumber \\
\bar{\delta}\, (m_{W_n}^2) &=& m_n^2\, \frac{15}{2}\, \frac{g_2^2}{16\pi^2}
\ln\frac{\Lambda^2}{\mu^2}, \nonumber \\
\bar{\delta}\, (m_{g_n}^2) &=& m_n^2 \,\frac{23}{2}\, \frac{g_3^2}{16\pi^2}
\ln\frac{\Lambda^2}{\mu^2}, \nonumber \\
\bar{\delta}\, (m_{H_n}^2) &=& m_n^2 \, \left(\frac{3}{2}\, g_2^2   
+ \frac{3}{4}\, g^{\prime 2} - \lambda_H \right)
\frac{1}{16\pi^2}\ln\frac{\Lambda^2}{\mu^2} + \overline{m}_H^2\ . 
\label{delta1}
\eear
Here $\lambda_H$ is the Higgs quartic coupling, ${\cal L}\supset -(\lambda_H/2)
(H^\dagger H)^2$, ($m_h=\sqrt{\lambda_H} v,\, v=246$ GeV), and $\overline{m}_H^2$
is the boundary mass term for the Higgs. The renormalization scale $\mu$
should be taken to be approximately the mass of the corresponding KK mode.
In the above formulae, we have not
included contributions from Yukawa couplings, which can be ignored
except for the top quark Yukawa coupling. Including the top Yukawa coupling
introduces no new corrections to the Higgs KK modes, but the KK modes of
the third generation $SU(2)$ doublet quark $Q_3$ and the $SU(2)$ singlet
$t$ receive additional corrections, 
\bear
\bar{\delta}_{h_t} m_{Q_{3n}} &=&m_n \left( -\frac{3}{4} \frac{h_t^2}{16\pi^2}\,
\ln \frac{\Lambda^2}{\mu^2}\right) \nonumber \\
\bar{\delta}_{h_t} m_{t_{n}} &=&m_n \left( -\frac{3}{2} \frac{h_t^2}{16\pi^2}\,
\ln \frac{\Lambda^2}{\mu^2}\right)\ .
\eear

The corrected masses for most of the KK modes can simply be
read off from equations~(\ref{delta0}) and (\ref{delta1}). However,
for certain fields there are also non-negligible tree-level 
contributions from electro-weak symmetry breaking, which introduce 
mixings among the states. This effect is important
for the ``photon'' and ``$Z$'', the two KK modes of the
top quark, the Higgs boson KK modes, and to a lesser extent 
for the bottom and tau KK modes.

\begin{figure}[tb]
\includegraphics[width=.48\textwidth]{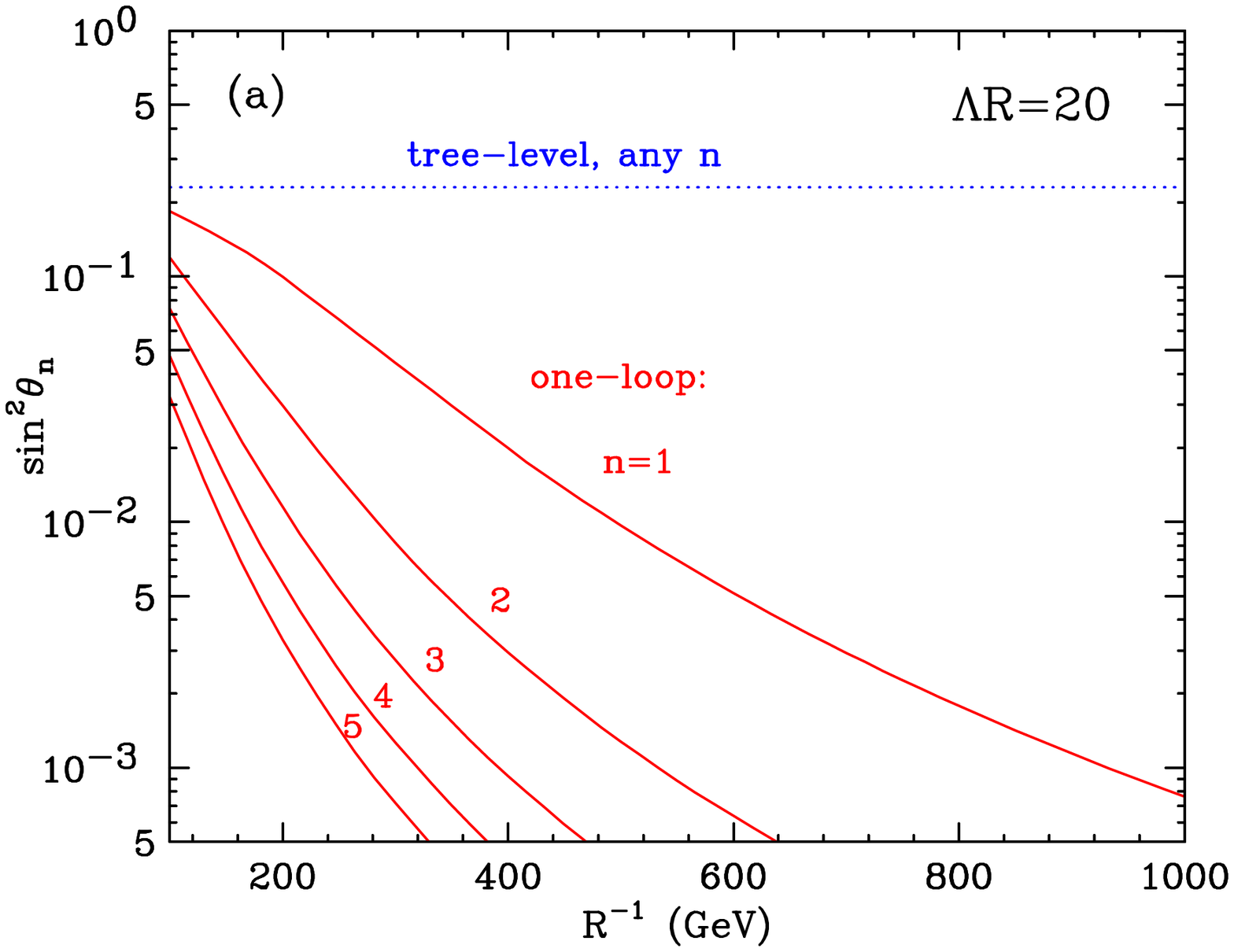}
\includegraphics[width=.48\textwidth]{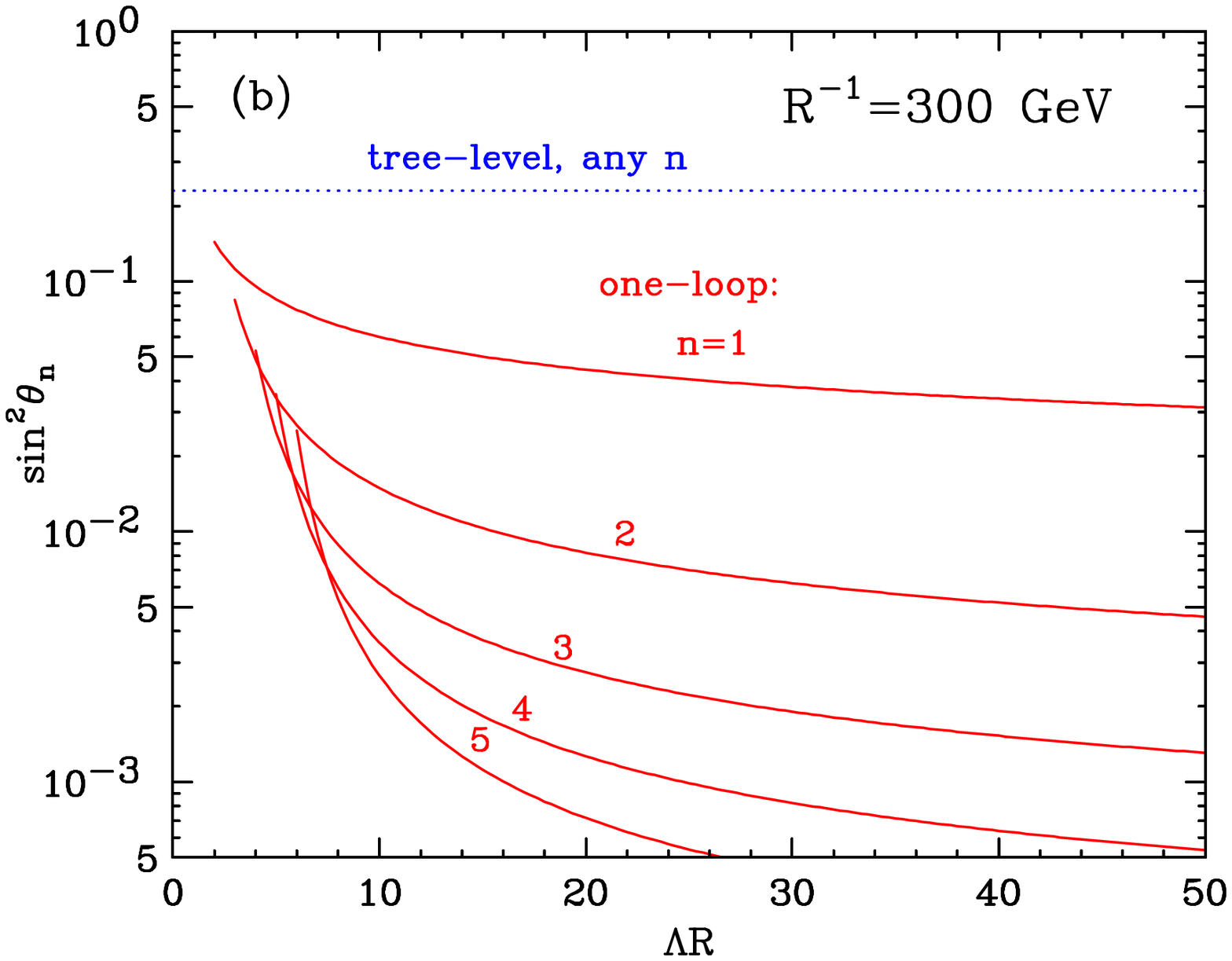}
\caption{\label{fig:mixing} 
Dependence of the ``Weinberg'' angle $\theta_n$ for the first few KK levels 
($n=1,2,...,5$) on 
(a) $R^{-1}$ for fixed $\Lambda R = 20$ and
(b) $\Lambda R $ for fixed $R^{-1} = 300$ GeV.
}
\end{figure}

The mass eigenstates and eigenvalues of the KK ``photons'' and 
``$Z$'s'' are obtained by diagonalizing their mass squared matrix.
In the $B_n$, $W^3_n$ basis it is 
\be
\left( 
\begin{array}{cc}
\frac{n^2}{R^2}+ \hat{\delta} m_{B_n}^2 + \frac{1}{4}g'^2 v^2 
& \frac{1}{4}g' g_2 v^2 \\
\frac{1}{4}g' g_2 v^2 & \frac{n^2}{R^2}+ \hat{\delta} m_{W_n}^2 +\frac{1}{4}g_2^2 v^2
\end{array}
\right),
\ee
where $\hat{\delta}$ represents the total one-loop correction,
including both bulk and boundary contributions.
Note that the mixing angle is different from the zero
mode Weinberg angle because of the corrections 
$\hat{\delta} m_{B_n}^2$ and $\hat{\delta} m_{W_n}^2$.
Fig.~\ref{fig:mixing} shows the dependence of the mixing angle $\theta_n$ 
for the n-th KK level on (a) $R^{-1}$ for fixed $\Lambda R = 20$; and
(b) $\Lambda R $ for fixed $R^{-1} = 300$ GeV.
For large $R^{-1}$ or $\Lambda R $, where the corrections become sizable,
the neutral gauge boson eigenstates become approximately pure $B_n$ and $W^3_n$.
%
%

Similarly, the eigenstates and eigenvalues of the KK fermions are obtained from
the corresponding mass matrices. For example, the mass matrix for the top
KK modes is
\be
\left(
\begin{array}{cc}
\frac{n}{R}+ \hat{\delta} m_{T_n} & m_t \\
m_t & -\frac{n}{R}-\hat{\delta} m_{t_n}
\end{array}
\right),
\ee
where $T_n$ and $t_n$ represent $SU(2)$ doublet quarks and singlet quarks 
respectively.

Finally we discuss the KK modes of the Higgs field. The KK modes 
of $W$ and $Z$ acquire their masses by ``eating'' linear combinations 
of the fifth component of the gauge fields and the Higgs KK modes. 
The orthogonal combinations remain physical scalar
particles. For $1/R\, \gg M_{W,Z}$, the longitudinal components
of the KK gauge bosons mostly come from $A_5$, and the physical scalars
are approximately the KK excitations of the Higgs field. There are
4 states at each KK level, $H^{\pm}_n,\, H^0_n,\, A^0_n$
(notice that $H^{\pm}_0$ and $A^0_0$ are just the usual Goldstone bosons in the SM).
Their corrected masses are given by
\bear
m_{H^0_n}^2 &\approx & m_n^2 + m_h^2 + \hat{\delta} m_{H_n}^2   \nonumber \\
m_{H^{\pm}_n}^2 &\approx & m_n^2 + M_W^2 + \hat{\delta} m_{H_n}^2   \nonumber \\
m_{A^0_n}^2 &\approx & m_n^2 + M_Z^2 + \hat{\delta} m_{H_n}^2\ . 
\eear

\begin{figure}[tb]
\includegraphics[width=.48\textwidth]{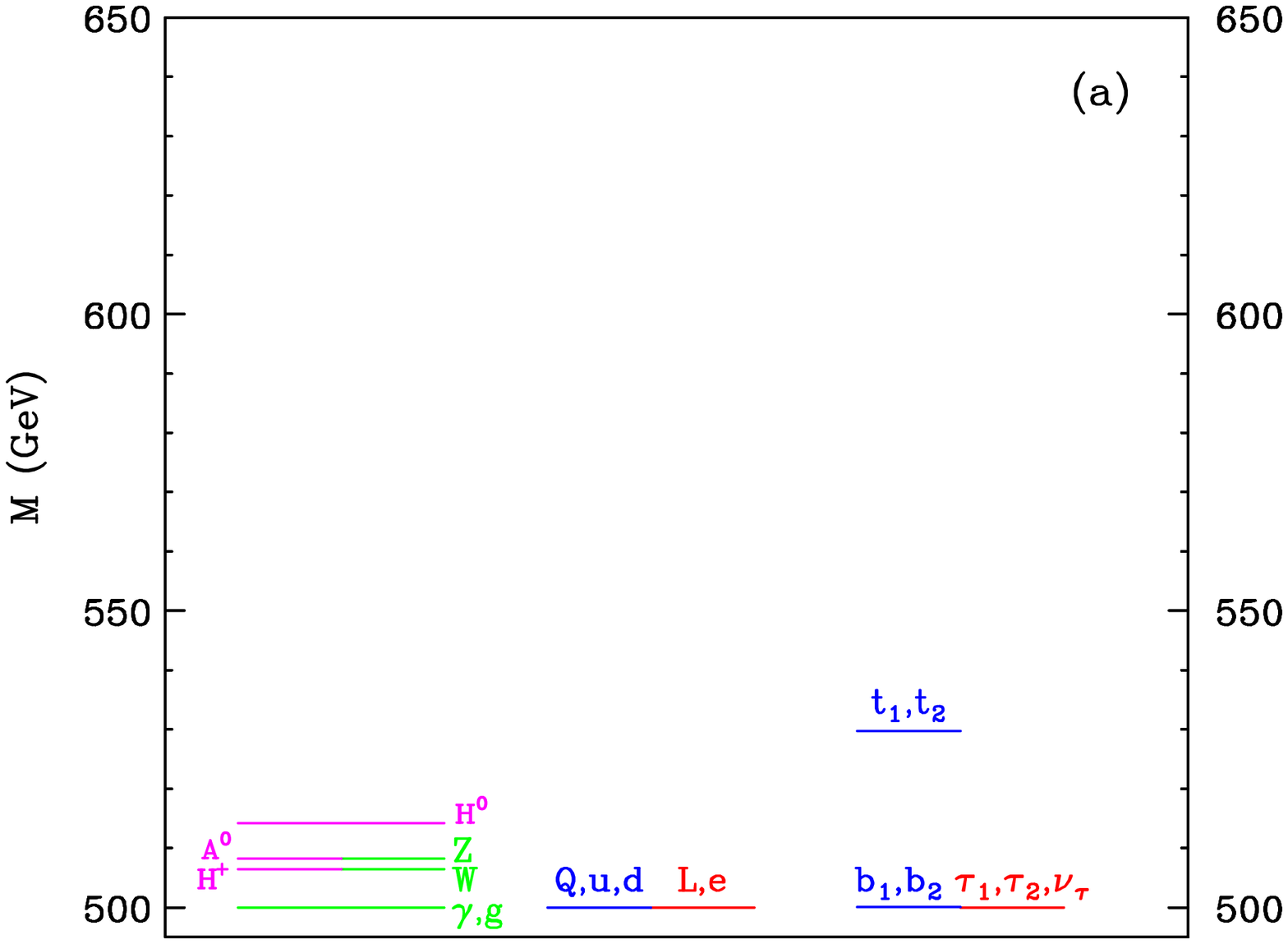}
\includegraphics[width=.48\textwidth]{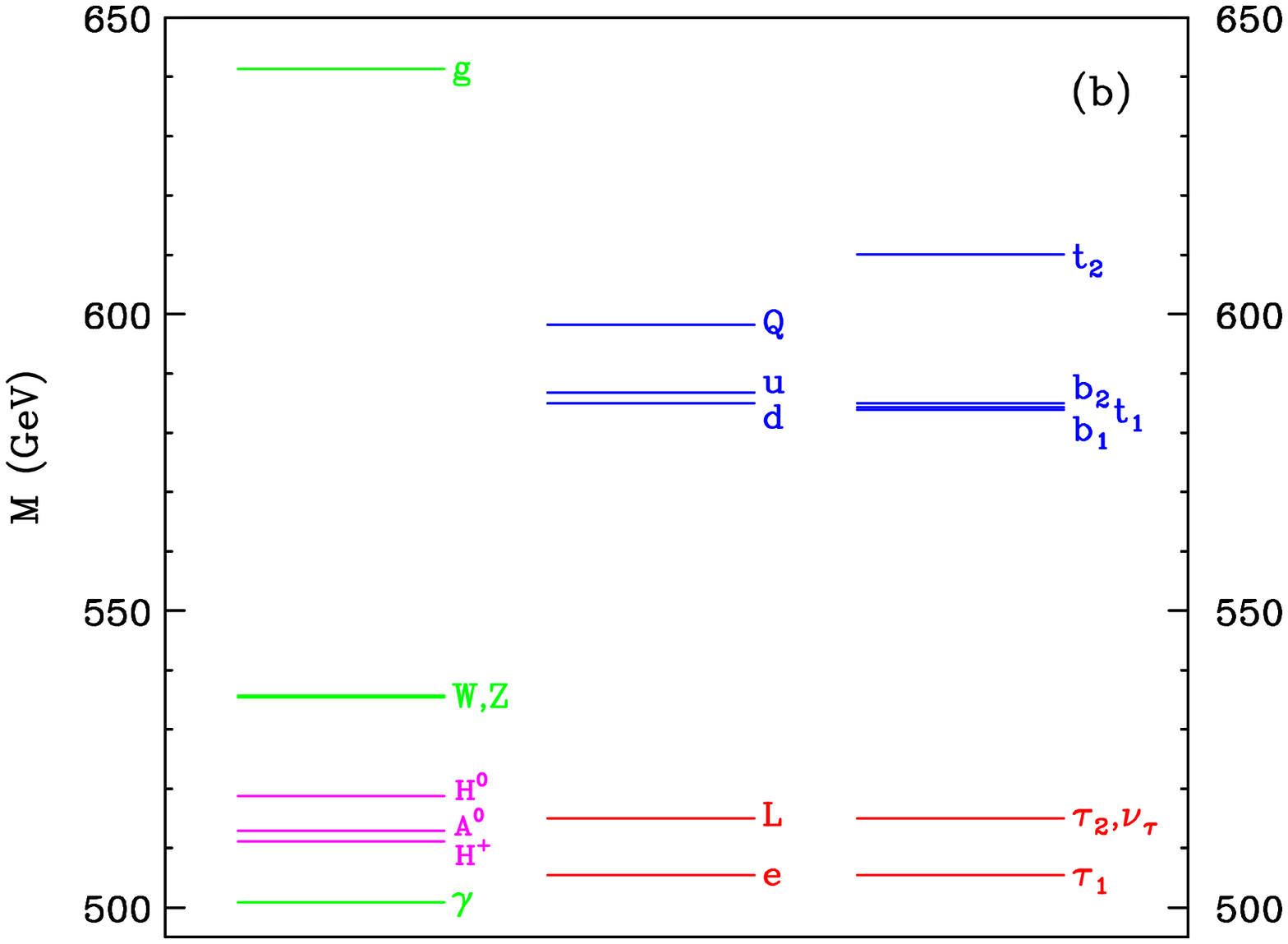}
\caption{\label{fig:spectrum} The spectrum of the first KK level
at (a) tree level and (b) one-loop, for 
$R^{-1}=500$ GeV, $\Lambda R = 20$, $m_h=120$ GeV,
$\overline{m}_H^2=0$, and
assuming vanishing boundary terms at the cut-off scale $\Lambda$.}
\end{figure}

In Fig.~\ref{fig:spectrum}, we show a sample spectrum for  
the first KK excitations of all Standard Model fields, both at tree-level (a)
and including the one-loop corrections (b).
%
%
We have fixed $R^{-1}=500$ GeV, $\Lambda R = 20$, $m_h=120$ GeV,
$\overline{m}_H^2=0$ and assumed vanishing boundary terms
at the cut-off scale $\Lambda$.
We see that the KK ``photon'' receives the smallest corrections
and is the lightest state at each KK level.
Unbroken KK parity $(-1)^{KK}$ implies that the lightest 
KK particle (LKP) at level one is stable. Hence the ``photon'' LKP
$\gamma_1$ provides an interesting dark matter candidate.
The corrections to the masses of the other first
level KK states are generally large enough that they will have
prompt cascade decays down to $\gamma_1$.\footnote{The first level 
graviton $G_1$ (or right-handed neutrino $N_1$ if the theory
includes right handed neutrinos $N_0$) could also be the LKP.
However, the decay lifetime of $\gamma_1$ to $G_1$ or 
$N_1$ would be comparable to cosmological scales. Therefore, 
$G_1$ and $N_1$ are
irrelevant for collider phenomenology but may have interesting
consequences for cosmology.}
Therefore KK production at colliders results in generic missing
energy signatures, similar to supersymmetric models with stable 
neutralino LSP. Collider searches for this scenario appear to
be rather challenging because of the KK mass degeneracy
and will be discussed in a separate publication~\cite{CMS}. 

\section{Conclusions} 

Loop corrections to the masses of Kaluza-Klein excitations can play
an important role in the phenomenology of extra dimensional theories.
This is because KK states of a given level are all nearly degenerate,
so that small corrections can determine which states decay and which
are stable. 

In this paper we computed the corrections to the masses of the 
KK excitations of gauge fields, scalars and spin-$\frac12$ 
fermions with arbitrary couplings in several extra-dimensional scenarios. 
Our results for one and two circular extra dimensions are presented
in Section 2 and Appendix~\ref{app:bulk}. They are
finite and cut-off independent as long as the cut-off is
5d Lorentz invariant and local. In Section 3 we extended our
results to the case of orbifolds $S^1/Z_2$ and $T^2/Z_2$. We
found divergences which introduce cut-off dependence.
The corresponding counter terms can be seen to be localized at
the fixed points of the orbifold.

In Section 4 we apply these results to the Standard Model in
extra dimensions and give explicit formulae for the corrected
masses of all KK excitations.
We hope that these results will be useful to practitioners of
the phenomenology of universal extra dimensions and other models
with Standard Model fields in the ``bulk'' (intriguing examples
are \cite{barbieri,ACG}).

\begin{acknowledgments}
We thank N.~Arkani-Hamed, A.~Cohen and B.~Dobrescu
for useful discussions. We also thank the Aspen Center for Physics
for hospitality during the initial stage of this work.
H.-C. C. is supported by the Department of Energy grant DE-FG02-90ER-40560.  
M.S. is supported in part by the Department of Energy under grant
number DE-FG02-91ER-40676. 
\end{acknowledgments}

\appendix

\section{One-loop bulk contributions}
\label{app:bulk}

In this Appendix we list the one-loop corrections to KK masses
from various diagrams with nonzero winding numbers. 

\begin{figure}[tb]
\begin{center}
{
\unitlength=1.0 pt
\SetScale{1.0}
\SetWidth{0.5}      
\normalsize    
\begin{picture}(100,100)(0,0)
\Photon(0.0,50.0)(30.0,50.0){2}{3}
\Photon(70.0,50.0)(100.0,50.0){2}{3}
\PhotonArc(50,50)(20,0,180){2}{8}
\PhotonArc(50,50)(20,180,360){2}{8}
\Text(50,80)[]{$A_\lambda$}
\Text(50,20)[]{$A_k$}
\Text(20,15)[]{(a)}
\end{picture} \
{} \qquad\allowbreak
\begin{picture}(100,100)(0,0)
\Photon(0.0,50.0)(30.0,50.0){2}{3}
\Photon(70.0,50.0)(100.0,50.0){2}{3}
\PhotonArc(50,50)(20,0,180){2}{8}
\PhotonArc(50,50)(20,180,360){2}{8}
\Text(50,80)[]{$A_\lambda$}
\Text(50,20)[]{$A_5$}
\Text(20,15)[]{(b)}
\end{picture} \
{} \qquad\allowbreak
\begin{picture}(100,100)(0,0)
\Photon(0.0,50.0)(30.0,50.0){2}{3}
\Photon(70.0,50.0)(100.0,50.0){2}{3}
\PhotonArc(50,50)(20,0,180){2}{8}
\PhotonArc(50,50)(20,180,360){2}{8}
\Text(50,80)[]{$A_5$}
\Text(50,20)[]{$A_5$}
\Text(20,15)[]{(c)}
\end{picture} \
{} \qquad\allowbreak
\begin{picture}(100,100)(0,0)
\Photon(0.0,50.0)(30.0,50.0){2}{3}
\Photon(70.0,50.0)(100.0,50.0){2}{3}
\DashCArc(50,50)(20,0,360){2}
\Text(50,80)[]{ghost}
\Text(20,15)[]{(d)}
\end{picture} \
{} \qquad\allowbreak
\begin{picture}(100,100)(0,0)
\Photon(0.0,30.0)(100.0,30.0){2}{11}
\PhotonArc(50,50)(20,0,360){3}{16}
\Text(50,80)[]{$A_\lambda$}
\Text(20,15)[]{(e)}
\end{picture} \
{} \qquad\allowbreak
\begin{picture}(100,100)(0,0)
\Photon(0.0,30.0)(100.0,30.0){2}{11}
\PhotonArc(50,50)(20,0,360){3}{16}
\Text(50,80)[]{$A_5$}
\Text(20,15)[]{(f)}
\end{picture} \
{} \qquad\allowbreak
\begin{picture}(100,100)(0,0)
\Photon(0.0,50.0)(30.0,50.0){2}{3}
\Photon(70.0,50.0)(100.0,50.0){2}{3}
\CArc(50,50)(20,0,360)
\Text(20,15)[]{(g)}
\end{picture} \
{} \qquad\allowbreak
\begin{picture}(100,100)(0,0)
\Photon(0.0,50.0)(30.0,50.0){2}{3}
\Photon(70.0,50.0)(100.0,50.0){2}{3}
\DashCArc(50,50)(20,0,360){3}
\Text(20,15)[]{(h)}
\end{picture} \
{} \qquad\allowbreak
\begin{picture}(100,100)(0,0)
\Photon(0.0,30.0)(100.0,30.0){2}{8}
\DashCArc(50,50)(20,0,360){3}
\Text(20,15)[]{(i)}
\end{picture} \
{} \qquad\allowbreak
}
\caption{\label{fig:bulk_v} One-loop diagrams for the gauge boson self energy,
(a) $A_\lambda -A_\kappa$ loop, (b) $A_\lambda -A_5$ loop,
(c) $A_5 -A_5$ loop, (d) ghost loop, (e) $A_\lambda$ loop, (f) $A_5$ loop,
(g) fermion loop,
(h) scalar-scalar loop, (i) scalar loop. 
}
\end{center}
\end{figure}
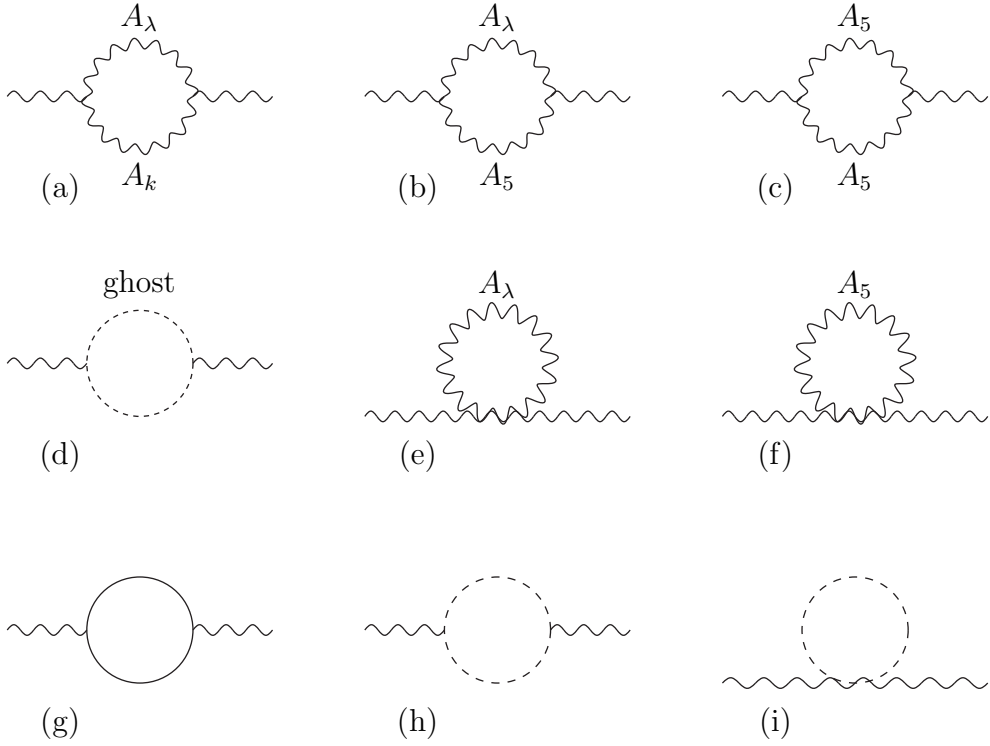

We consider one extra dimension compactified on a circle $S^1$ with radius $R$.
The various one-loop diagrams for the gauge boson self energy are shown in
Fig.~\ref{fig:bulk_v}.
%
%
The contributions from nonzero winding numbers to the zero mode and nonzero modes
in the Feynman-'t~Hooft gauge are listed in Table~\ref{tab:bulk_v}.
\begin{table}
\caption{\label{tab:bulk_v} The contributions from the diagrams in 
Fig.~\ref{fig:bulk_v}(a)-(i). 
All these terms are multiplied $\frac{g^2 \zeta (3)}{16\pi^4 R^2}$. 
For the scalar loops in (h) and (i), the results are for each real component.
}
\begin{ruledtabular}
\begin{tabular}{ccc}
Diagram & Nonzero mode & Zero mode \\
\hline
(a) & $C(G)$ & $-\frac{9}{2}C(G)$ \\
(b) & $-2C(G)$ & $C(G)$ \\
(c) & $0$ & $-C(G)$ \\
(d) & $0$ & $\frac{1}{2}C(G)$ \\
(e) & $3C(G)$ & $3C(G)$ \\
(f) & $C(G)$ & $C(G)$ \\
(g) & $-4T(r_f)$ & $0$ \\
(h) & $0$ & $-T(r_s)$ \\
(i) & $T(r_s)$ & $T(r_s)$ \\
\end{tabular}
\end{ruledtabular}
\end{table}
After summing over all diagrams, we find that the total contribution
to the zero mode is $0$, and the contribution to nonzero modes is
\be
\delta m^2_{V_{KK}}=\frac{g^2 \zeta(3)}{16\pi^4 R^2} \left(3 C(G) +
\sum_{\rm real\; scalars} T(r_s) -4\sum_{\rm fermions} T(r_f)\right).
\ee

The one-loop contribution to the fermion self energy is also obtained
easily. For the example of QED,
\be
\Sigma = - 3e^2 \int d\alpha \sum_{k_5} \int \frac{d^4 k_E}{(2\pi)^4}
\frac{\alpha (\not{p} + i \gamma_5 p_5)  + i\gamma_5 k'_5}{[k_E^2+k^{\prime 2}_5
- \alpha (1-\alpha) (p^2-p_5^2)]^2},
\ee
where $k'_5= k_5-\alpha p_5$. The term proportional to $k'_5$
vanishes after Poisson resummation. The remainder is a function
of $\not{p} +i\gamma_5 p_5$ and therefore does not contribute to
KK mode masses. Similar arguments apply to all other fermion self
energy diagrams.

Scalar masses are not protected by symmetries, and they can receive
power-divergent contributions. However, we can use the same method to
isolate the finite contributions from loops with nonzero
winding numbers. We find that these finite corrections
are the same for zero mode and nonzero modes in the leading order
for $m_0 \ll 1/R$. They are both given by
\be
\frac{\zeta(3)}{16\pi^4 R^2} \left(
4\, g^2\, T(r) +   \sum_{\rm real\; scalars} \frac{\lambda}{2}
- \sum_{\rm 4-comp\; fermions} 4 h_f^2 \right)
\label{scalar}
\ee
and can be absorbed into the overall mass term. At the lowest order, 
there is no relative correction between zero mode and nonzero mode.

One can also generalize to more extra dimensions. For example, we consider 
two extra dimensions compactified on a square torus with radius $R$ for both
dimensions. The result is very similar to the one extra dimension case, 
except that the factor
\be
\zeta(3)=\sum_{n=1}^{\infty} \frac{1}{n^3} \approx 1.202
\ee
in the 5-dimensional formulae is replaced by
\be
\frac{1}{\pi} \sum_{m,n\in Z}^{m^2+n^2\neq 0}
 \frac{1}{(m^2+n^2)^2}
=\frac{4}{\pi} \left(\zeta (4) + \sum_{m=1}^{\infty} \sum_{n=1}^{\infty}
\frac{1}{(m^2+n^2)^2} \right)
\equiv \frac{4}{\pi} \left(\zeta(4)+\Delta \right)
\approx \frac{4}{\pi}\times 1.506,
\ee
and one has to include the $A_6$ loop, which contributes like a real adjoint
scalar. There is also an extra adjoint scalar at each KK level coming from 
a linear combination of $A_5$ and $A_6$, which is not eaten by the KK gauge
bosons. The correction to the KK mode masses of the gauge boson and the extra
adjoint scalar are the same and are both given by
\be
\delta m^2_{V_{KK}}(6D)=\frac{g^2 (\zeta(4)+\Delta)}{4\pi^5 R^2} \left(4 C(G) +
\sum_{\rm real\; scalars} T(r_s) -\sum_{\rm 4-comp\, fermions} 4T(r_f)\right).
\ee

\section{One-loop boundary contributions}
\label{app:boundary}

In this Appendix, we list the one-loop contributions to the boundary
terms for gauge fields, fermions, and scalars for the $S^1/Z_2$ orbifold
compactification. The results for the case of a two dimensional orbifold 
$T^2/Z_2$ are briefly discussed at the end.

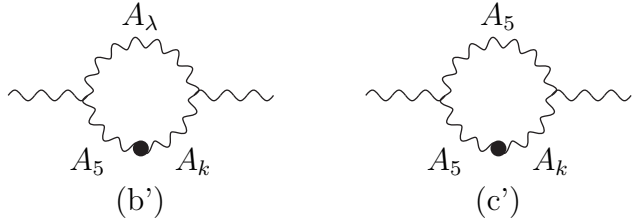
\begin{figure}[tb]
\begin{center}
{
\unitlength=1.0 pt
\SetScale{1.0}
\SetWidth{0.5}      
\normalsize    
\begin{picture}(100,100)(0,0)
\Photon(0.0,50.0)(30.0,50.0){2}{3}
\Photon(70.0,50.0)(100.0,50.0){2}{3}
\PhotonArc(50,50)(20,0,180){2}{8}
\PhotonArc(50,50)(20,180,360){2}{8}
\Text(50,80)[]{$A_\lambda$}
\Text(30,25)[]{$A_5$}
\Text(70,25)[]{$A_k$}
\Vertex(50,30){3}
\Text(50,10)[]{(b')}
\end{picture} \
{} \qquad\allowbreak
\begin{picture}(100,100)(0,0)
\Photon(0.0,50.0)(30.0,50.0){2}{3}
\Photon(70.0,50.0)(100.0,50.0){2}{3}
\PhotonArc(50,50)(20,0,180){2}{8}
\PhotonArc(50,50)(20,180,360){2}{8}
\Text(50,80)[]{$A_5$}
\Text(30,25)[]{$A_5$}
\Text(70,25)[]{$A_k$}
\Vertex(50,30){3}
\Text(50,10)[]{(c')}
\end{picture} \
{} \qquad\allowbreak
}
\caption{\label{fig:xi_gauge} 
Additional divergent contributions to the gauge boson self energy
in the $\xi$ gauge where $A_\mu$ and $A_5$ do not decouple.
(b') $A_\lambda-A_5,A_{\kappa}$ loop, (c') $A_5-A_5,A_{\kappa}$ loop.
}
\end{center}
\end{figure}

The one-loop diagrams for the gauge boson self energy are shown in
Fig.~\ref{fig:bulk_v}. We keep only logarithmically divergent
contributions to the boundary terms. They can be written as
\be
\overline{\Pi}_{\mu\nu} 
=\frac{g^2}{64\pi^2}\, \ln \frac{\Lambda^2}{\mu^2}
\Bigg\{g_{\mu\nu}\, p^2\, a_1 - p_\mu p_\nu\, a_2
+ g_{\mu\nu}\, \frac{p_5^2+p^{\prime 2}_5}{2}\, a_3 \Bigg\},
\quad \left(\text{for } p'_5=p_5+\frac{2n}{R}\right).
\ee
In Table~\ref{tab:boundary_v}, we list $a_1,\, a_2,\, a_3$ in the
$\xi$ gauge (using the gauge fixing of the 5 dimensional generalized 
Lorentz gauge condition). In this gauge, $A_\mu$ and $A_5$ do not decouple
for $\xi \neq 1$, so there are additional divergent diagrams shown in
Fig.~\ref{fig:xi_gauge}.
%
%
\begin{table}
\caption{\label{tab:boundary_v} The contributions to $a_1,\, a_2,\, a_3$
from the diagrams in 
Fig.~\ref{fig:bulk_v}(a)-(i) and Fig.~\ref{fig:xi_gauge}(b'),(c').
There is no contribution from fermions at one-loop due to the cancellation
between the $Z_2$ even and odd fermion components. 
For the scalar loops in (h) and (i), the upper (lower) sign is for 
the $Z_2$ even (odd) scalar, and 
the results are for each real component.
}
\begin{ruledtabular}
\begin{tabular}{cccc}
Diagram & $a_1$ & $a_2$ & $a_3$ \\
\hline
(a) & $\left[\frac{19}{6} - (\xi-1)\right] C(G)$ & 
      $\left[\frac{11}{3} - (\xi-1)\right] C(G)$ &
      $\left[\frac{9}{2}+\frac{9}{4}(\xi-1) \right] C(G)$ \\
(b) & $0$ & $0$ & $\left[ 3+\frac{3}{4}(\xi-1)\right] C(G)$ \\
(b') & $0$ & $0$ & $\frac{3}{2}(\xi-1) C(G)$ \\
(c) & $\frac{1}{3} C(G)$ & $\frac{1}{3}C(G)$ & 
      $\left[-1+(\xi-1)\right] C(G)$ \\
(c') & $0$ & $0$ & $-2(\xi-1) C(G)$ \\
(d) & $\frac{1}{6} C(G)$ & $-\frac{1}{3}C(G)$ & $-\frac{1}{2} C(G)$ \\
(e) & $C(G)$ & $0$ & $\left[-3-\frac{3}{2}(\xi-1)\right] C(G)$ \\
(f) & $C(G)$ & $0$ & $\left[1-(\xi-1)\right] C(G)$ \\
(g) & $0$ & $0$ & $0$ \\
(h) & $\mp \frac{1}{3} T(r_s)$ & $\mp \frac{1}{3} T(r_s)$ & $\pm T(r_s)$ \\
(i) & $0$ & $0$ & $\mp T(r_s)$ \\
\end{tabular}
\end{ruledtabular}
\end{table}
Adding all contributions together, we obtain
\bear
\overline{\Pi}_{\mu\nu}
&=&
\frac{g^2}{64\pi^2}\, \ln \frac{\Lambda^2}{\mu^2}
\Bigg\{(g_{\mu\nu}\, p^2- p_\mu p_\nu)\Bigg[
\bigg(\frac{11}{3}-(\xi-1)\bigg) \, C(G) 
- \frac{1}{3}
\sum_{\rm real\; scalars}\bigg(T(r)_{\rm even} -T(r)_{\rm odd}\bigg)\Bigg]
\nonumber \\
&+& g_{\mu\nu}\, 
\frac{(p_5^2+p^{\prime 2}_5)}{2}\,\bigg(4+(\xi-1)\bigg)\, C(G) \Bigg\},
\quad \quad \quad \left(\text{for } p'_5=p_5+\frac{2n}{R}\right).
\eear

The correction to the squared mass of the $n$-th mode KK gauge boson can be
obtained from the term proportional to $g_{\mu\nu}$, by setting 
$p^2= p^2_5 =p^{\prime 2}_5= m_n^2=n^2/R^2$, and multiplying
by the wave function normalization factor $(\sqrt{2})^2$,
\be
\bar{\delta} m_{V_n}^2= m_n^2\, 
\frac{g^2}{32\pi^2}\, \ln \frac{\Lambda^2}{\mu^2}
\left[ \frac{23}{3}\, C(G) -\frac{1}{3} 
\sum_{\rm real\; scalars}\bigg(T(r)_{\rm even} -T(r)_{\rm odd}\bigg)\right].
\label{gauge1}
\ee
One can see that the result is gauge independent.

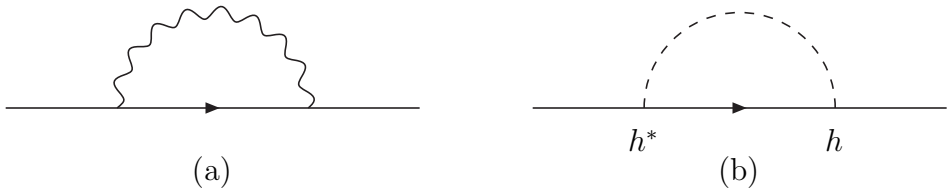
\begin{figure}[tb]
\begin{center}
{
\unitlength=1.2 pt
\SetScale{1.2}
\SetWidth{0.5}      
\normalsize    
\begin{picture}(140,100)(0,30)
\ArrowLine(5.0,50.0)(135.0,50.0)
\PhotonArc(70,50)(30,0,180){2}{9}
\Text(70,30)[]{(a)}
\end{picture}
{} \qquad\allowbreak
\begin{picture}(140,100)(0,30)
\ArrowLine(5.0,50.0)(135.0,50.0)
\DashCArc(70,50)(30,0,180){3}
\Text(70,30)[]{(b)}
\Text(40,40)[]{$h^*$}
\Text(100,40)[]{$h$}
\end{picture}
{} \qquad\allowbreak
}
\caption{\label{fig:boundary_f} One-loop diagrams for the fermion self energy,
(a) gauge boson loop, (b) scalar boson loop. }
\end{center}
\end{figure}

The one-loop fermion self energy diagrams are shown in 
Fig.~\ref{fig:boundary_f}.
%
%
Keeping only the logarithmically divergent contributions, we can write
\be
\overline{\Sigma} 
=\frac{1}{64\pi^2}\, \ln \frac{\Lambda^2}{\mu^2}
\Bigg[ \not{p} \frac{1\pm\gamma_5}{2}\, b_1 - \bigg(i p_5 \frac{1\pm\gamma_5}{2}
- i p'_5 \frac{1\mp\gamma_5}{2}\bigg)\, b_2 \Bigg],
\quad \left(\text{for } p'_5=p_5+\frac{2n}{R}\right).
\ee
The contributions to $b_1,\, b_2$ are listed in Table~\ref{tab:boundary_f}.
\begin{table}
\caption{\label{tab:boundary_f} The contributions to $b_1,\, b_2$ from 
the diagrams in Fig.~\ref{fig:boundary_f}(a),(b). 
$C(r)$ is defined $C(r)\delta_{ij}= \sum_a
T^a_{ik} T^a_{kj}$ ($= (N^2-1)/(2N)$ for the fundamental representation
of $SU(N)$ gauge group).
The upper (lower) sign in (b) is for $Z_2$ even (odd) scalars.
}
\begin{ruledtabular}
\begin{tabular}{ccc}
Diagram & $b_1$ & $b_2$ \\
\hline
(a) & $\left[-1-2(\xi-1)\right]\, g^2\, C(r)$ & 
      $\left[5+(\xi-1)\right]\, g^2\, C(r)$ \\
(b) & $\mp h^2$ & $\mp h^2$ \\
\end{tabular}
\end{ruledtabular}
\end{table}
The correction to the fermion KK mode mass is given by
\be
\bar{\delta} m_{f_n} = m_n\, \frac{1}{64\pi^2} \ln\frac{\Lambda^2}{\mu^2}
\left[ 9\,C(r)\,g^2- \sum_{\rm even\, scalars} 3\, h_+^2
+\sum_{\rm odd\, scalars} 3\, h_-^2\right].
\ee

\begin{figure}[tb]
\begin{center}
{
\unitlength=1.0 pt
\SetScale{1.0}
\SetWidth{0.5}      
\normalsize    
\begin{picture}(100,100)(0,0)
\DashLine(0.0,40.0)(100.0,40.0){3}
\PhotonArc(50,40)(25,0,180){2}{9}
\Text(50,75)[]{$A_\lambda$}
\Text(20,15)[]{(a)}
\end{picture} \
{} \qquad\allowbreak
\begin{picture}(100,100)(0,0)
\DashLine(0.0,40.0)(100.0,40.0){3}
\PhotonArc(50,40)(25,0,180){2}{9}
\Text(50,75)[]{$A_5$}
\Text(20,15)[]{(b)}
\end{picture} \
{} \qquad\allowbreak
\begin{picture}(100,100)(0,0)
\DashLine(0.0,30.0)(100.0,30.0){3}
\PhotonArc(50,50)(20,0,360){3}{15}
\Text(50,80)[]{$A_\lambda$}
\Text(20,15)[]{(c)}
\end{picture} \
{} \qquad\allowbreak
\begin{picture}(100,100)(0,0)
\DashLine(0.0,30.0)(100.0,30.0){3}
\PhotonArc(50,50)(20,0,360){3}{15}
\Text(50,80)[]{$A_5$}
\Text(20,15)[]{(d)}
\end{picture} \
{} \qquad\allowbreak
\begin{picture}(100,100)(0,0)
\DashLine(0.0,50.0)(30.0,50.0){3}
\DashLine(70.0,50.0)(100.0,50.0){3}
\CArc(50,50)(20,0,360)
\Text(20,15)[]{(e)}
\Text(22,42)[]{$h^*$}
\Text(75,42)[]{$h$}
\end{picture} \
{} \qquad\allowbreak
\begin{picture}(100,100)(0,0)
\DashLine(0.0,30.0)(100.0,30.0){3}
\DashCArc(50,50)(20,0,360){3}
\Text(20,15)[]{(f)}
\Text(50,22)[]{$\lambda$}
\end{picture} \
{} \qquad\allowbreak
}
\caption{\label{fig:boundary_s} One-loop diagrams for the scalar boson 
self energy,
(a) $A_\lambda$-scalar loop, (b) $A_5$-scalar loop,
(c) $A_\lambda$ loop, (d) $A_5$ loop,
(e) fermion loop,
(f) scalar loop. 
}
\end{center}
\end{figure}
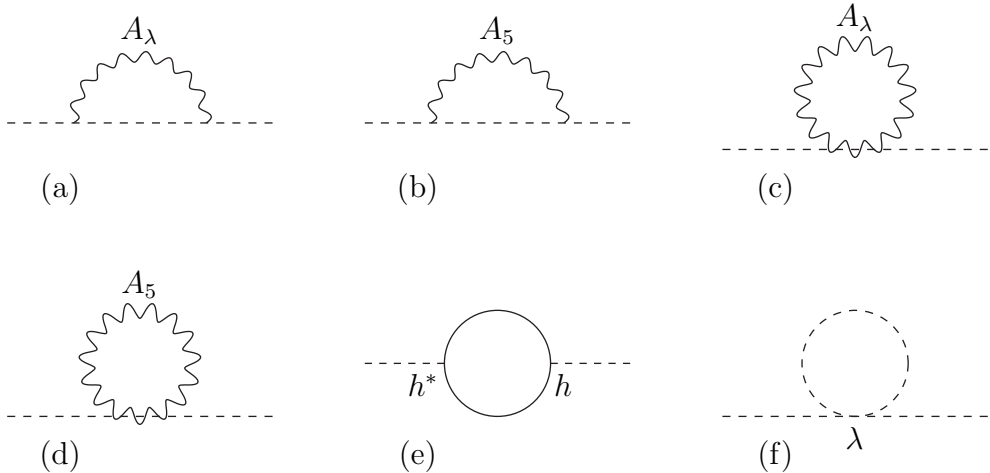

A $Z_2$ even scalar can receive power-divergent contributions to
both the bulk mass term and the boundary mass term. 
We need to fine tune these mass terms to have a light scalar.
The boundary mass term causes mixing among KK modes and we
need to re-diagonalize the mass matrix to find the eigenstates if
it is large.
The possibility of a light scalar arising because of cancellation 
between the bulk mass and the boundary mass may be interesting, but
will not be considered here.
Instead, we assume that both the bulk mass and the boundary mass are
tuned to be much smaller than the compactification scale, so that
we can treat the boundary mass term as a small perturbation and
ignore the higher order mixing effects. The boundary mass term
can be written as
\be
\frac{L}{2}\bigg( \delta(x_5) +\delta(x_5-L)\bigg) \overline{m}^2\, 
\Phi^\dagger \Phi.
\ee
Using the KK decomposition, (\ref{kkdecomp}), we find that the contribution
to the zero mode is $\overline{m}^2$, while to the nonzero mode is
$2\,\overline{m}^2$, due to the normalization factor $\sqrt{2}$ at the 
boundaries. Therefore, the nonzero KK modes receive a correction 
$\overline{m}^2$ relative to the zero mode from the boundary mass term,
(ignoring a weak scale dependence
due to the wave function renormalization.)
We can also calculate the correction due to the boundary kinetic terms.
The one-loop diagrams for the scalar self energy are shown in 
Fig.~\ref{fig:boundary_s}.
%
%
They can be written as
\be
\frac{1}{64\pi^2}\, \ln \frac{\Lambda^2}{\mu^2}
\Bigg\{p^2\, c_1 
+\frac{p_5^2+p^{\prime 2}_5}{2}\, c_2 \Bigg\},
\quad \quad \left(\text{for } p'_5=p_5+\frac{2n}{R}\right).
\ee
The coefficients $c_1, \, c_2$ (in the Feynman gauge) are given 
in Table~\ref{tab:boundary_se}.
\begin{table}
\caption{\label{tab:boundary_se} The contributions to $c_1,\, c_2$ 
(in Feynman gauge) from 
the diagrams in Fig.~\ref{fig:boundary_s}(a)-(f). 
The upper (lower) sign in (f) is for an $Z_2$ even (odd) scalar in the loop.
}
\begin{ruledtabular}
\begin{tabular}{ccc}
Diagram & $c_1$ & $c_2$ \\
\hline
(a) & $4 g^2 T(r)$ & $2g^2 T(r)$ \\
(b) & $0$ & $3g^2 T(r)$ \\
(c) & $0$ & $-4g^2 T(r)$ \\
(d) & $0$ & $g^2 T(r)$ \\
(e) & $0$ & $0$ \\
(f) & $0$ & $\mp \frac{\lambda}{2}$ \\
\end{tabular}
\end{ruledtabular}
\end{table}
Including the normalization factor $(\sqrt{2})^2$, we have the correction
to the KK modes of an even scalar,
\be
\bar{\delta} m^2_{S+_{n}} = \overline{m}^2 + m_n^2 \frac{1}{32\pi^2}
\ln \frac{\Lambda^2}{\mu^2} \left[ 6g^2 T(r) - \sum_{\rm even\, scalars}
\frac{\lambda_{++}}{2} +\sum_{\rm odd\, scalars} \frac{\lambda_{+-}}{2}
\right],
\ee
where the sum is over real components.

For an odd scalar, there is no boundary mass term. The correction
comes only from boundary kinetic terms, 
\be
\frac{1}{64\pi^2}\, \ln \frac{\Lambda^2}{\mu^2} \,
p_5 p^{\prime}_5\, d_1,
\quad \quad \left(\text{for } p'_5=p_5+\frac{2n}{R}\right).
\ee
The coefficients $d_1$ from one-loop diagrams are listed in 
Table~\ref{tab:boundary_so}.
\begin{table}
\caption{\label{tab:boundary_so} The contributions to $d_1$ 
(in Feynman gauge) from 
the diagrams in Fig.~\ref{fig:boundary_s}(a)-(f). 
The upper (lower) sign in (f) is for an $Z_2$ even (odd) scalar in the loop.
}
\begin{ruledtabular}
\begin{tabular}{cc}
Diagram & $d_1$  \\
\hline
(a) & $0$ \\
(b) & $5g^2 T(r)$ \\
(c) & $4g^2 T(r)$ \\
(d) & $-g^2 T(r)$ \\
(e) & $0$  \\
(f) & $\pm \frac{\lambda}{2}$ \\
\end{tabular}
\end{ruledtabular}
\end{table}
The total correction to the KK modes of an odd scalar KK is
\be
\bar{\delta} m^2_{S-_{n}} =  m_n^2 \frac{1}{32\pi^2}
\ln \frac{\Lambda^2}{\mu^2} \left[ 9g^2 T(r) + \sum_{\rm even\, scalars}
\frac{\lambda_{+-}}{2} -\sum_{\rm odd\, scalars} \frac{\lambda_{--}}{2}
\right].
\ee

Finally, we briefly describe the results for 2 extra dimensions
compactified on a $T^2/Z_2$ orbifold, with a square torus $T^2$ of
radius $R$ for each side. The $Z_2$ is a $180^o$ rotation in the $x_5,\, x_6$
plane, which flips the signs of both $x_5$ and $x_6$. The gauge components
$A_5,\, A_6$ are odd under $Z_2$ while $A_\mu,\, \mu=0,\,1,\,2,\,3$ are even.
There will be induced terms localized at the orbifold fixed points
$(x_5,x_6)=(0,\,0),\,(0,\, \pi R),\, (\pi R,\, 0),\, (\pi R,\,\pi R)$,
which break 6-dimensional Lorentz invariance.

The KK states are labeled by a pair of KK numbers, $(n_1,\, n_2)$, with
$(n_1,\, n_2)$ and $(-n_1,\, -n_2)$ identified. There are KK parities
associated with each KK number.
The results are similar to the 5-dimensional case on $S^1/Z_2$, except
that we need to include the extra $A_6$ component, which contributes
like an odd adjoint real scalar. We have
\bear
\bar{\delta} m_{V_{(n_1, n_2)}}^2 &=& m_{(n_1,n_2)}^2\, 
\frac{g^2}{32\pi^2}\, \ln \frac{\Lambda^2}{\mu^2}
\left[ 8\, C(G) -\frac{1}{3} 
\sum_{\rm real\; scalars}\bigg(T(r)_{\rm even} -T(r)_{\rm odd}\bigg)\right],
\\
\bar{\delta} m_{f_{(n_1,n_2)}} &=& m_{(n_1,n_2)}\, \frac{1}{64\pi^2} 
\ln\frac{\Lambda^2}{\mu^2}
\left[ 12\,C(r)\,g^2- \sum_{\rm even\, scalars} 3\, h_+^2
+\sum_{\rm odd\, scalars} 3\, h_-^2\right],
\\
\bar{\delta} m^2_{S+_{(n_1,n_2)}} &=& \overline{m}^2 + m_{(n_1,n_2)}^2 \,
\frac{1}{32\pi^2}
\ln \frac{\Lambda^2}{\mu^2} \left[ 7g^2 T(r) - \sum_{\rm even\, scalars}
\frac{\lambda_{++}}{2} +\sum_{\rm odd\, scalars} \frac{\lambda_{+-}}{2}
\right],
\\
\bar{\delta} m^2_{S-_{(n_1,n_2)}} &=&  m_{(n_1,n_2)}^2 \frac{1}{32\pi^2}
\ln \frac{\Lambda^2}{\mu^2} \left[ 8g^2 T(r) + \sum_{\rm even\, scalars}
\frac{\lambda_{+-}}{2} -\sum_{\rm odd\, scalars} \frac{\lambda_{--}}{2}
\right].
\eear
In addition, there are also KK states corresponding to the linear
combination  of $A_5$ and $A_6$ which is not eaten by the KK gauge
boson. These KK states are odd adjoint scalars.
Their corrections are just like the odd adjoint scalars'
\be
\bar{\delta} m_{P_{(n_1, n_2)}}^2= m_{(n_1,n_2)}^2\, 
\frac{g^2}{32\pi^2}\, \ln \frac{\Lambda^2}{\mu^2}
\left[ 8\, C(G) +
\sum_{\rm real\; scalars}\bigg(T(r)_{\rm even} -T(r)_{\rm odd}\bigg)\right].
\ee

\section{KK number violating couplings}
\label{app:vertex}

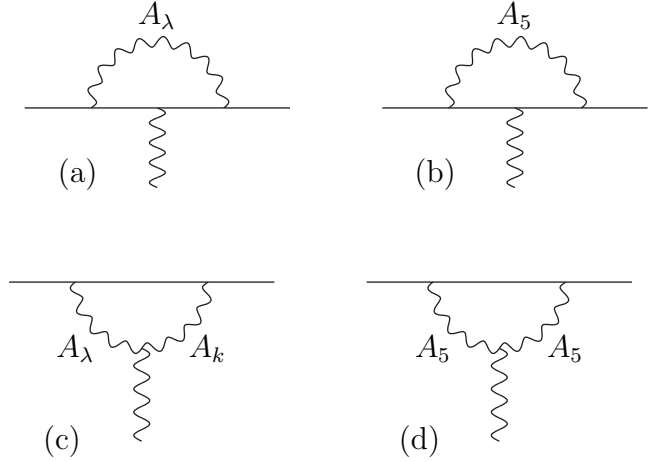
\begin{figure}[tb]
\begin{center}
{
\unitlength=1.0 pt
\SetScale{1.0}
\SetWidth{0.5}      
\normalsize    
\begin{picture}(100,100)(0,0)
\Line(0.0,40.0)(100.0,40.0)
\PhotonArc(50,40)(25,0,180){2}{9}
\Photon(50,40)(50,10){3}{4}
\Text(50,75)[]{$A_\lambda$}
\Text(20,15)[]{(a)}
\end{picture} \
{} \qquad\allowbreak
\begin{picture}(100,100)(0,0)
\Line(0.0,40.0)(100.0,40.0)
\PhotonArc(50,40)(25,0,180){2}{9}
\Photon(50,40)(50,10){3}{4}
\Text(50,75)[]{$A_5$}
\Text(20,15)[]{(b)}
\end{picture} \\
{} \qquad\allowbreak
\begin{picture}(100,100)(0,0)
\Line(0.0,75.0)(100.0,75.0)
\PhotonArc(50,75)(25,180,360){2}{9}
\Photon(50,50)(50,15){3}{4}
\Text(25,50)[]{$A_\lambda$}
\Text(75,50)[]{$A_k$}
\Text(20,15)[]{(c)}
\end{picture} \
{} \qquad\allowbreak
\begin{picture}(100,100)(0,0)
\Line(0.0,75.0)(100.0,75.0)
\PhotonArc(50,75)(25,180,360){2}{9}
\Photon(50,50)(50,15){3}{4}
\Text(25,50)[]{$A_5$}
\Text(75,50)[]{$A_5$}
\Text(20,15)[]{(d)}
\end{picture} \
{} \qquad\allowbreak
}
\caption{\label{fig:vertex_loop} One-loop diagrams for the fermion-gauge boson
interaction
(a) $A_\lambda$--fermion--fermion loop, (b) $A_5$--fermion--fermion loop,
(c) $A_\lambda -A_\kappa$--fermion loop, 
(d) $A_5 -A_5$--fermion loop,  
}
\end{center}
\end{figure}

In this Appendix we discuss the KK number violating couplings in an orbifold
compactification.  Using the example of one extra dimension 
compactified on $S^1/Z_2$, we consider the KK number violating couplings
between the fermion and the gauge field. Fig.~\ref{fig:vertex_loop}
shows the one-loop vertex corrections for the fermion gauge interactions.
%
%
The contributions to the KK number violating interaction are logarithmically
divergent. They can be written as
\be
\bar{\delta}{\cal L} \supset -\frac{L}{2}\bigg( \delta (x_5)+\delta (x_5-L) \bigg)\,
f_1\,\frac{g^2}{64\pi^2}\, \ln \frac{\Lambda^2}{\mu^2} \,
g\,\overline{\psi} \gamma^\mu T^a P_+ \psi\, A_\mu^a ,
\ee 
where $P_+ =P_R$ or $P_L$ is the projection on the $Z_2$ even fermions.
The coefficients $f_1$ from the diagrams in Fig.~\ref{fig:vertex_loop}
are listed in Table~\ref{tab:vertex}. 
\begin{table}
\caption{\label{tab:vertex} The contributions to $f_1$ from 
the diagrams in Fig.~\ref{fig:vertex_loop}(a)-(d). 
}
\begin{ruledtabular}
\begin{tabular}{cc}
Diagram & $f_1$  \\
\hline
(a) & $[2C(r)-C(G)] [1+(\xi-1)]$ \\
(b) & $-C(r)+\frac{1}{2} C(G)$ \\
(c) & $C(G) [3+\frac{3}{2}(\xi-1)]$ \\
(d) & $-\frac{1}{2}C(G)$ \\
\end{tabular}
\end{ruledtabular}
\end{table}
Summing over them gives
\be
f_1({\rm total})= C(r) \bigg[1+2(\xi-1)\bigg] 
+ C(G)\bigg[2+\frac{1}{2}(\xi-1)\bigg].
\ee

To obtain the couplings among the physical mass eigenstates, we need to 
include the KK number violating mass and kinetic mixing effects on
the external legs, since they are also one-loop effects.
The (4-dimensional) kinetic mixing needs to be treated with some care.
We illustrate this with a simple example of two real scalars,
$\phi_p,\, \phi_q$, with masses $m_p< m_q$, and a small kinetic mixing
proportional to $\epsilon$.
\be
{\cal L} = \frac{1}{2} \partial_\mu \phi_p \partial^\mu \phi_p
+ \epsilon \partial_\mu \phi_p \partial^\mu \phi_q
+\frac{1}{2} \partial_\mu \phi_q \partial^\mu \phi_q
- \frac{1}{2}  m_p^2 \phi_p^2 -\frac{1}{2} m_q^2 \phi_q^2.
\ee
We will only work in the leading order of $\epsilon$. First, we re-define
$\phi_p$ to absorb the mixing term,
\be
\begin{array}{l}
\phi'_p = \phi_p + \epsilon \phi_q \\
\phi'_q \approx \phi_q
\end{array}
\quad\quad {\rm or} \quad\quad
\begin{array}{l}
\phi_p \approx \phi'_p - \epsilon \phi'_q \\
\phi_q \approx \phi'_q
\end{array}.
\label{kin_mixing}
\ee
In terms of $\phi'_p,\, \phi'_q$, the mass terms become
\be
- \frac{1}{2}  m_p^2 \phi^{\prime 2}_p 
+ \epsilon m_p^2 \phi'_p \phi'_q
-\frac{1}{2} m_q^2 \phi^{\prime 2}_q.
\ee
Now we can diagonalize the mass matrix by a rotation between $\phi'_p$ and 
$\phi'_q$. The physical eigenstates $\phi''_p$ and $\phi''_q$ are given
approximately by
\bear
\phi''_p &\approx & \phi'_p + \frac{\epsilon m_p^2}{m_q^2-m_p^2} \phi'_q
\approx \phi_p + \frac{\epsilon m_q^2}{m_q^2-m_p^2} \phi_q 
\nonumber \\
\phi''_q &\approx & \phi'_q - \frac{\epsilon m_p^2}{m_q^2-m_p^2} \phi'_p
\approx \phi_q - \frac{\epsilon m_p^2}{m_q^2-m_p^2} \phi_p.
\eear
In particular, if one of them is massless, $m_p=0$, the relation
between the physical states and the original states is simply given
by eq.(\ref{kin_mixing}).

%
%
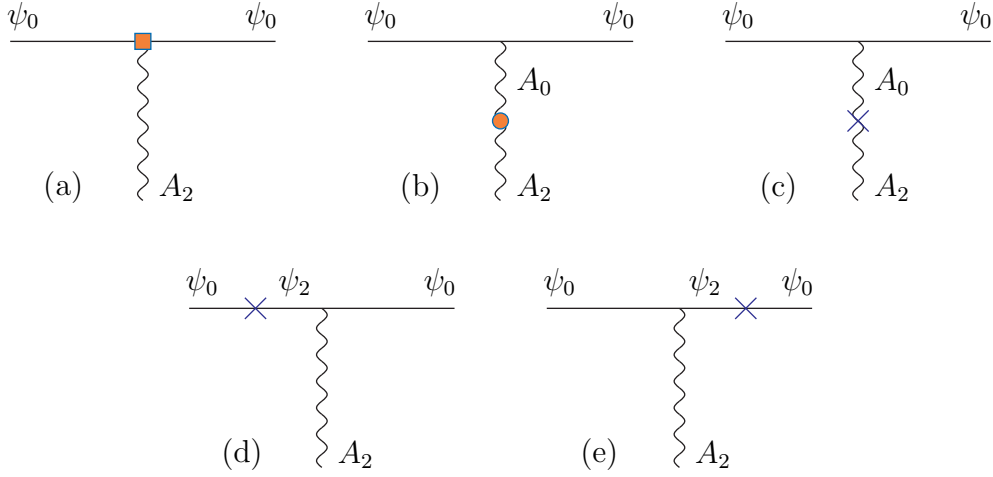
\begin{figure}[tb]
\begin{center}
{
\unitlength=1.0 pt
\SetScale{1.0}
\SetWidth{0.5}      
\normalsize    
\begin{picture}(100,100)(0,0)
\Line(0.0,70.0)(100.0,70.0)
\Photon(50,70)(50,10){2}{6}
\Text( 5,80)[]{$\psi_0$}
\Text(95,80)[]{$\psi_0$}
\Text(63,15)[]{$A_2$}
\CBoxc(50,70)(6,6){NavyBlue}{Orange}
\Text(20,15)[]{(a)}
\end{picture} \
{} \qquad\allowbreak
\begin{picture}(100,100)(0,0)
\Line(0.0,70.0)(100.0,70.0)
\Photon(50,70)(50,10){2}{6}
\Text( 5,80)[]{$\psi_0$}
\Text(95,80)[]{$\psi_0$}
\CCirc(50,40){3}{NavyBlue}{Orange}
\Text(63,15)[]{$A_2$}
\Text(63,55)[]{$A_0$}
\Text(20,15)[]{(b)}
\end{picture} \
{} \qquad\allowbreak
\begin{picture}(100,100)(0,0)
\Line(0.0,70.0)(100.0,70.0)
\Photon(50,70)(50,10){2}{6}
\Text( 5,80)[]{$\psi_0$}
\Text(95,80)[]{$\psi_0$}
\SetColor{Blue}
\Line(54,44)(46,36)
\Line(54,36)(46,44)
\SetColor{Black}
\Text(63,15)[]{$A_2$}
\Text(63,55)[]{$A_0$}
\Text(20,15)[]{(c)}
\end{picture} \
{} \qquad\allowbreak
\begin{picture}(100,100)(0,0)
\Line(0.0,70.0)(100.0,70.0)
\Photon(50,70)(50,10){2}{6}
\Text( 5,80)[]{$\psi_0$}
\Text(40,80)[]{$\psi_2$}
\Text(95,80)[]{$\psi_0$}
\SetColor{Blue}
\Line(29,74)(21,66)
\Line(29,66)(21,74)
\SetColor{Black}
\Text(63,15)[]{$A_2$}
\Text(20,15)[]{(d)}
\end{picture} \
{} \qquad\allowbreak
\begin{picture}(100,100)(0,0)
\Line(0.0,70.0)(100.0,70.0)
\Photon(50,70)(50,10){2}{6}
\Text( 5,80)[]{$\psi_0$}
\Text(60,80)[]{$\psi_2$}
\Text(95,80)[]{$\psi_0$}
\SetColor{Blue}
\Line(79,74)(71,66)
\Line(79,66)(71,74)
\SetColor{Black}
\Text(63,15)[]{$A_2$}
\Text(20,15)[]{(e)}
\end{picture} \
{} \qquad\allowbreak
}
\caption{\label{fig:f0f0A2} The KK number violating coupling for
$\overline{\psi}_{0} \gamma^\mu T^a P_+ \psi_0 A^a_{2\mu}$. The dot
represents the kinetic mixing and the cross represents the mass mixing.
The contributions from various diagrams are  $\sqrt{2} g \frac{g^2}{16\pi^2} 
\ln \frac{\Lambda^2}{\mu^2} \times$ 
(a) One-loop vertex: $\{C(r)[1+2(\xi-1)]+2C(G)[2+\frac{1}{2}(\xi-1)]\}$,
(b) $A_2$(external)--$A_0$ kinetic mixing: $\{[\frac{11}{3}-(\xi-1)]C(G)-
\frac{1}{3} \sum_{\rm real\, scalars} (T(r_+)-T(r_-))\}$,
(c) $A_2-A_0$ mass mixing: $[2+\frac{1}{2}(\xi-1)]C(G)$,
(d), (e) $\psi_0-\psi_2$ mass mixing: , $\{-[5+(\xi-1)C(r)\}\times 2$. 
}
\end{center}
\end{figure}
As an example, we compute the coupling between the mass eigenstates of a
second (or $2n$-th) KK mode gauge boson and two zero mode fermions. 
The contributions are shown in Fig.~\ref{fig:f0f0A2}.
Combining all contributions we obtain the $\overline{\psi}_0- \psi_0-A_2$
interaction vertex to be
\bear
&&(-i\gamma^\mu g T^a P_+)\sqrt{2}\, 
\frac{g^2}{64\pi^2}\,\ln\frac{\Lambda^2}{\mu^2}\,
\left[ \frac{23}{3}\, C(G) -\frac{1}{3} 
\sum_{\rm real\; scalars}\bigg(T(r)_{\rm even} -T(r)_{\rm odd}\bigg)
-9\, C(r) \right] \nonumber \\
&=&(-i\gamma^\mu g T^a P_+) \frac{\sqrt{2}}{2} \left[
\frac{\bar{\delta}(m_{A_2}^2)}{m_{2}^2}- 2\,\frac{\bar{\delta} (m_{f_2})}{m_{2}}
\right].
\eear
It is not too surprising that it is related to the mass corrections from 
the boundary terms. The $\sqrt{2}$ factor comes from the normalization
of the KK mode at the boundaries.

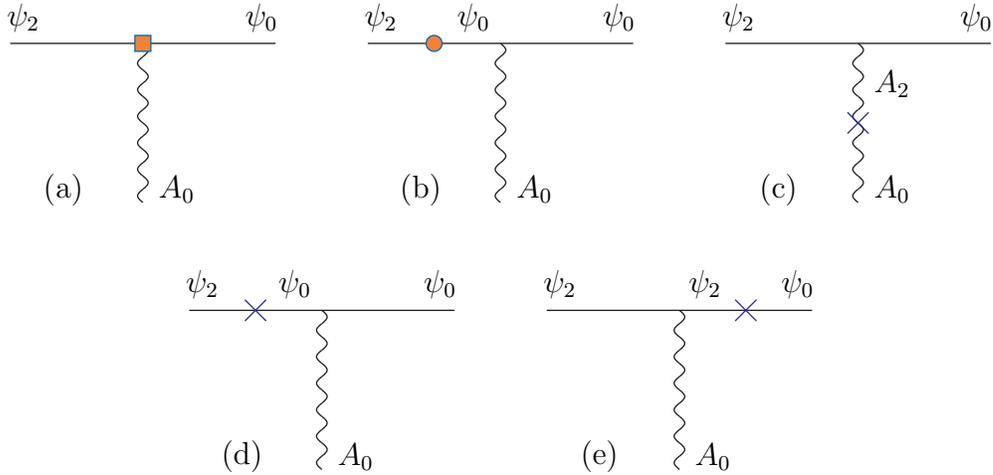
\begin{figure}[tb]
\begin{center}
{
\unitlength=1.0 pt
\SetScale{1.0}
\SetWidth{0.5}      
\normalsize    
\begin{picture}(100,100)(0,0)
\Line(0.0,70.0)(100.0,70.0)
\Photon(50,70)(50,10){2}{6}
\Text( 5,80)[]{$\psi_2$}
\Text(95,80)[]{$\psi_0$}
\Text(63,15)[]{$A_0$}
\CBoxc(50,70)(6,6){NavyBlue}{Orange}
\Text(20,15)[]{(a)}
\end{picture} \
{} \qquad\allowbreak
\begin{picture}(100,100)(0,0)
\Line(0.0,70.0)(100.0,70.0)
\Photon(50,70)(50,10){2}{6}
\CCirc(25,70){3}{NavyBlue}{Orange}
\Text( 5,80)[]{$\psi_2$}
\Text(40,80)[]{$\psi_0$}
\Text(95,80)[]{$\psi_0$}
\Text(63,15)[]{$A_0$}
\Text(20,15)[]{(b)}
\end{picture} \
{} \qquad\allowbreak
\begin{picture}(100,100)(0,0)
\Line(0.0,70.0)(100.0,70.0)
\Photon(50,70)(50,10){2}{6}
\Text( 5,80)[]{$\psi_2$}
\Text(95,80)[]{$\psi_0$}
\SetColor{Blue}
\Line(54,44)(46,36)
\Line(54,36)(46,44)
\SetColor{Black}
\Text(63,15)[]{$A_0$}
\Text(63,55)[]{$A_2$}
\Text(20,15)[]{(c)}
\end{picture} \
{} \qquad\allowbreak
\begin{picture}(100,100)(0,0)
\Line(0.0,70.0)(100.0,70.0)
\Photon(50,70)(50,10){2}{6}
\Text( 5,80)[]{$\psi_2$}
\Text(40,80)[]{$\psi_0$}
\Text(95,80)[]{$\psi_0$}
\SetColor{Blue}
\Line(29,74)(21,66)
\Line(29,66)(21,74)
\SetColor{Black}
\Text(63,15)[]{$A_0$}
\Text(20,15)[]{(d)}
\end{picture} \
{} \qquad\allowbreak
\begin{picture}(100,100)(0,0)
\Line(0.0,70.0)(100.0,70.0)
\Photon(50,70)(50,10){2}{6}
\Text( 5,80)[]{$\psi_2$}
\Text(60,80)[]{$\psi_2$}
\Text(95,80)[]{$\psi_0$}
\SetColor{Blue}
\Line(79,74)(71,66)
\Line(79,66)(71,74)
\SetColor{Black}
\Text(63,15)[]{$A_0$}
\Text(20,15)[]{(e)}
\end{picture} \
{} \qquad\allowbreak
}
\caption{\label{fig:f2f0A0} The KK number violating coupling for
$\overline{\psi}_{2} \gamma^\mu T^a P_+ \psi_0 A^a_{0\mu}$. The dot
represents kinetic mixing and the cross represents mass mixing.
The contributions from various diagrams are  $\sqrt{2} g \frac{g^2}{16\pi^2} 
\ln \frac{\Lambda^2}{\mu^2} \times$ 
(a) One-loop vertex: $\{C(r)[1+2(\xi-1)]+2C(G)[2+\frac{1}{2}(\xi-1)]\}$,
(b) $\psi_2$(external)--$\psi_0$ kinetic mixing: $\{-[1+2(\xi-1)]\}$,
(c) $A_2-A_0$ mass mixing: $\{-[2+\frac{1}{2}(\xi-1)]C(G)\}$,
(d) $\psi_2$(external)--$\psi_0$ mass mixing: $[5+(\xi-1)]C(r)$,
(e) $\psi_0$(external)--$\psi_2$ mass mixing: $\{-[5+(\xi-1)]C(r)\}$. 
}
\end{center}
\end{figure}
One can also check the KK number violating couplings involving the
zero mode gauge boson, e.g., 
$\overline{\psi}_{2} \gamma^\mu T^a P_+ \psi_0 A^a_{0\mu}$ 
(Fig.~\ref{fig:f2f0A0}).
We find that they vanish as required by gauge invariance.\footnote{However, 
there can be higher dimensional operators such as 
$\overline{\psi}_{2} \sigma^{\mu\nu} T^a P_+ \psi_0 F^a_{0\mu\nu}$.}


\newpage

\begin{thebibliography}{99}

\bibitem{Cheng:1998hc}
H.-C.~Cheng, B.~A.~Dobrescu and K.~T.~Matchev,
``Generic and chiral extensions of the supersymmetric standard model,''
Nucl.\ Phys.\ B {\bf 543}, 47 (1999)
[arXiv:hep-ph/9811316].

\bibitem{Appelquist:2001nn}
T.~Appelquist, H.-C.~Cheng and B.~A.~Dobrescu,
``Bounds on universal extra dimensions,''
Phys.\ Rev.\ D {\bf 64}, 035002 (2001)
[arXiv:hep-ph/0012100].

\bibitem{Georgi:2001ks}
H.~Georgi, A.~K.~Grant and G.~Hailu,
``Brane couplings from bulk loops,''
Phys.\ Lett.\ B {\bf 506}, 207 (2001)
[arXiv:hep-ph/0012379].

\bibitem{Hosotani}
Y.~Hosotani,
``Dynamical Mass Generation By Compact Extra Dimensions,''
Phys.\ Lett.\ B {\bf 126}, 309 (1983);
%
H.~Hatanaka, T.~Inami and C.~S.~Lim,
``The gauge hierarchy problem and higher dimensional gauge theories,''
Mod.\ Phys.\ Lett.\ A {\bf 13}, 2601 (1998),
[arXiv:hep-th/9805067];
%
I.~Antoniadis, K.~Benakli and M.~Quiros,
``Finite Higgs mass without supersymmetry,''
arXiv:hep-th/0108005;
G.~V.~Gersdorff, N.~Irges and M.~Quiros,
``Bulk and brane radiative effects in gauge theories on orbifolds,''
arXiv:hep-th/0204223.

\bibitem{Contino:2001nj}
R.~Contino, L.~Pilo, R.~Rattazzi and A.~Strumia,
``Graviton loops and brane observables,''
JHEP {\bf 0106}, 005 (2001)
[arXiv:hep-ph/0103104].

\bibitem{Arkani-Hamed:2000hv}
N.~Arkani-Hamed, H.-C.~Cheng, B.~A.~Dobrescu and L.~J.~Hall,
``Self-breaking of the standard model gauge symmetry,''
Phys.\ Rev.\ D {\bf 62}, 096006 (2000)
[arXiv:hep-ph/0006238].

\bibitem{Dobrescu:2001ae}
B.~A.~Dobrescu and E.~Poppitz,
``Number of fermion generations derived from anomaly cancellation,''
Phys.\ Rev.\ Lett.\  {\bf 87}, 031801 (2001)
[arXiv:hep-ph/0102010].

\bibitem{Appelquist:2001mj}
T.~Appelquist, B.~A.~Dobrescu, E.~Ponton and H.~U.~Yee,
``Proton stability in six dimensions,''
Phys.\ Rev.\ Lett.\  {\bf 87}, 181802 (2001)
[arXiv:hep-ph/0107056].

\bibitem{Agashe:2001xt}
K.~Agashe, N.~G.~Deshpande and G.~H.~Wu,
``Universal extra dimensions and b $\to$ s gamma,''
Phys.\ Lett.\ B {\bf 514}, 309 (2001)
[arXiv:hep-ph/0105084].

\bibitem{Appelquist:2001jz}
T.~Appelquist and B.~A.~Dobrescu,
``Universal extra dimensions and the muon magnetic moment,''
Phys.\ Lett.\ B {\bf 516}, 85 (2001)
[arXiv:hep-ph/0106140].

\bibitem{Rizzo:2001sd}
T.~G.~Rizzo,
``Probes of universal extra dimensions at colliders,''
Phys.\ Rev.\ D {\bf 64}, 095010 (2001)
[arXiv:hep-ph/0106336].

\bibitem{Macesanu:2002db}
C.~Macesanu, C.~D.~McMullen and S.~Nandi,
``Collider implications of universal extra dimensions,''
arXiv:hep-ph/0201300;

\bibitem{CMS}
H.-C.~Cheng, K.~T.~Matchev and M.~Schmaltz,
to appear.

\bibitem{barbieri}
R.~Barbieri, L.~J.~Hall and Y.~Nomura,
``A constrained standard model from a compact extra dimension,''
Phys.\ Rev.\ D {\bf 63}, 105007 (2001)
[arXiv:hep-ph/0011311].

\bibitem{ACG}
N.~Arkani-Hamed, A.~G.~Cohen and H.~Georgi,
``Electroweak symmetry breaking from dimensional deconstruction,''
Phys.\ Lett.\ B {\bf 513}, 232 (2001)
[arXiv:hep-ph/0105239].


\end{thebibliography}
\end{document}